# On thought experiments: Mach and Einstein (Part I)

Eren Simsek**Abstract**

This paper will attempt to show that the Doppler principle was of major relevance to Einstein for the genesis of the special theory of relativity. For this purpose, I focus primarily on Einstein's speech "How I created the theory of relativity" (1922) and his work "On the Electrodynamics of moving bodies" (1905). His speech has already been studied by many researchers, however the focus of previous research was on whether and how the Michelson experiment influenced Einstein in the development of the special theory of relativity. The following paper will discuss the historical and social background of his speech in Kyoto, and provide analysis of Einstein's thought experiment in connection with the Michelson experiment. In this context, I will argue that the work of Ernst Mach and Woldemar Voigt had a significant impact on Einstein.## 1. Introduction

*Ichi-go ichi-e*

(Japanese Idiom)

Einstein's speech "How I created the theory of relativity," which he spontaneously held at the request of Kitaro Nishida in 1922 at Kyoto University in Japan, is interesting and unique in many ways.[i] Among other things, there are no existing notes of Einstein related to this speech and yet the content is preserved through a transcript (by Jun Ishiwara) – but only in Japanese. The speech has already been studied by many researchers (Pais 1982, Stachel 1982, Holton 1988, Itagaki 1999, Abiko 2000, Dongen 2009), with a focus on the Michelson experiment mentioned by Einstein in his speech. To clarify the question on the role of the Michelson experiment in the genesis of the special theory of relativity, various translations were written from Japanese (Ogawa 1979, Ukawa quoted in Stachel 1982, Ono 1983, Haubold and Yasui 1986, Itagaki 1999, Abiko 2000) some of them contradict one and other in their relevant text passages. The contradictory translations, led me to ask Hisaki Hashi, a Japanese native speaker with expertise in the philosophy of science for advice. She explained why East Asian texts are difficult to translate into Western languages.[ii] Since my work in English is intended for a wide audience, I offer the translation of *The Collected Papers of Albert Einstein* to the reader for my analysis. Hisaki Hashi considers that both translations (*The Collected Papers of Albert Einstein* and Haubold and Yasui 1986) are contextually correct. Personally, I prefer the German translation from Haubold and Yasui 1986.

The Einstein research has examined the speech in Kyoto, but to my knowledge has not provided enough historical context. The central thesis of this work is to present a historical reconstruction of the aforementioned speech, which led me to six theses:

(1) The relevance of the Doppler effect for the genesis of the theory of special relativity

> Important Einstein researchers such as John Stachel (Stachel 1982) and John Norton (Norton 2004, 2011) have presented good arguments, for the role the emission theory played in the genesis of special relativity. However, according to Einstein´s (later) statement, this was not the case (see Einstein´s thoughts on emission theory in Shankland 1963, pp. 49, 56). In this work I will try to offer an alternative to this 'dilemma':

> The Doppler effect significantly impacted Einstein in the creation of the special theory of relativity.[iii] This is the starting point of my investigation on the genesis of special relativity, as well as the history of the Doppler Effect and their relation to each other.

(2) The reconstruction of the Kyoto speech by Einstein

Using the source material, I reconstruct the content of the speech, as historically accurate as possible. In my estimation, Einstein´s explanation of how he created the special theory of relativity is in accordance with the history of science. His procedural approach is comprehensible from a scientific standpoint – especially if one gives the Doppler effect an important role in the emergence of the special relativity.

(3) Einstein´s thought experiments and their relation to Mach

From this perspective, in which the Doppler effect is credited with a significant role, it may be possible to better assess Ernst Mach´s influence on Einstein in relation to the emergence of the special relativity. Since E. Mach can be considered as one of the pioneers in the theory of thought experiments, I will focus on Einstein´s thought experiments and try to show that Mach´s work could have played a direct or indirect role.

(4) Mach´s influence on the special theory of relativity

There are arguments that Einstein could have known Mach´s work on the Doppler effect, and the special relativity was in Mach´s line of thought.

(5) Historical social background of the Kyoto speech by Einstein

Einstein´s repeated reference to Mach in his speech (1922) becomes particularly clear when one includes the historical social background. Namely, one can understand his speech as a kind of "defense", in which he questions Mach´s critical comments of the theory of relativity, which appeared in 1921.

(A short explanation of this case: Ernst Mach died tragically after decades of severe illness in 1916, shortly before the theory of general relativity was completed. Five years after Mach´s passing, his book *The Principles of Physical Optics* was published posthumously. In the preface of this book, Ernst Mach clearly states that he does not understand himself as a forerunner of the theory of relativity and will criticize this theory in the second volume of his book.

Such a second volume never appeared. Mach research proved the foreword to have been faked, by Ernst Mach's cocaine-addicted son Ludwig Mach, who was probably hoping, to receive financial gain from the Nazis (Wolters 1987). This incident led to two victims in this history of science: primarily Ernst Mach - because his scientific and philosophical reputation was discredited for decades following the publication. The second victim was Einstein, who experienced an attack from his mentor from the grave, even though Einstein thought very highly of Mach and had defended him for years. Einstein never knew, that the preface was not written from Mach. This work pays homage and is dedicated to them (Mach and Einstein).)

(6) On thought experiments: Mach and Einstein

Ernst Mach is considered a pioneer in the theory of thought experiments. Research in this field has focused on seeing Mach as a kind of sensationalist/positivist (see for example Sorensen 1998, p. 4; Brown 2011, p. 154) and his interpretations were seen in this light. From this point of view, it is difficult to understand the parallel of thoughts between Einstein and Mach. Finally, I will present the analysis of Rudolf Haller, who explained Mach´s concepts in a way, which will be utilized in the following paper to offer new insights on Mach´s influence on Einstein and the theory of relativity.

The, above mentioned, thesis partly depend on each other (for example, the 3 and 4, etc.). In general, I hope that the content of my work can make a small contribution, to Mach research, to Einstein research and thus to Mach Einstein research.

## 2. The historical and social background of Einstein´s speech

"Thanking you again with all my heart for your kind letter, I remain your admiring student,

A. Einstein"

(Einstein to Mach 1909)[iv]

The background of how this speech came to be (from Einstein) and how it was published can be found *The Collected Papers of Albert Einstein, Vol. 13, pp. 624 ff.* / (CPAE 13; _ ; pp. 624 ff.)[1]. The historical and social background I present should serve as a supplement to better understand the content of the speech and thus the motivation and intention of Einstein.

It is well known that Planck, in his strongly polemical writing, had declared Ernst Mach a false prophet[v], which became famous as the Mach-Planck debate (see Wolters 2019, §28.4). Mach, who at the time was not only sick[vi] but also elderly, had severe difficulties complaining. He had already mentioned Planck´s work in the second preface to his *Mechanics*. Of course, it should not go unmentioned that Planck himself was a Machian in his youth (see Blackmore and Hentschel 1985, p. 59, fn. 2).[vii] Mach was known not to be a prophet; neither right nor wrong – on the contrary, he had seen himself in the service of the Enlightenment.[viii] He was, inter alia, an outstanding experimental physicist who recognized the weaknesses of (Newtonian) mechanics and suggested solutions. However, the attack of Planck asked an important question besides his subject-relevant criticism: To what extent were Mach´s criticisms and concepts trend-setting for modern physics?

---

[1] I will continue to refer to *The Collected Papers of Albert Einstein* with CPAE X; Y; Z, where CPAE is *The Collected Papers of Albert Einstein*; X is the volume number; Y is the document number, if missing then left empty "_" – and Z is the page number of the English translation supplement. (I use the citation format from Janssen and Renn (2015) and add the page number to enable the reader to find the passages easier.)

The reaction to this question was the Mach-Einstein coalition[ix] (now partially forgotten). This Mach-Einstein coalition, as I call it, is well documented by Einstein´s letters to Mach (see Wolters 1987, pp. 148 ff.), by Mach´s endorsement of Einstein´s theory of relativity as a progression of his own ideas (Wolters 2019, §28.5), and Einstein's constant reference to Mach in his work on (general) relativity.

Nevertheless, the image of Mach in public is a completely different (see Wolters 2019, §28.2). Just one example is given here briefly to illustrate the situation: Ten years (1926) after E. Mach´s death, a bust was erected in his honor in Rathauspark (Vienna) (fig. 1)[x].

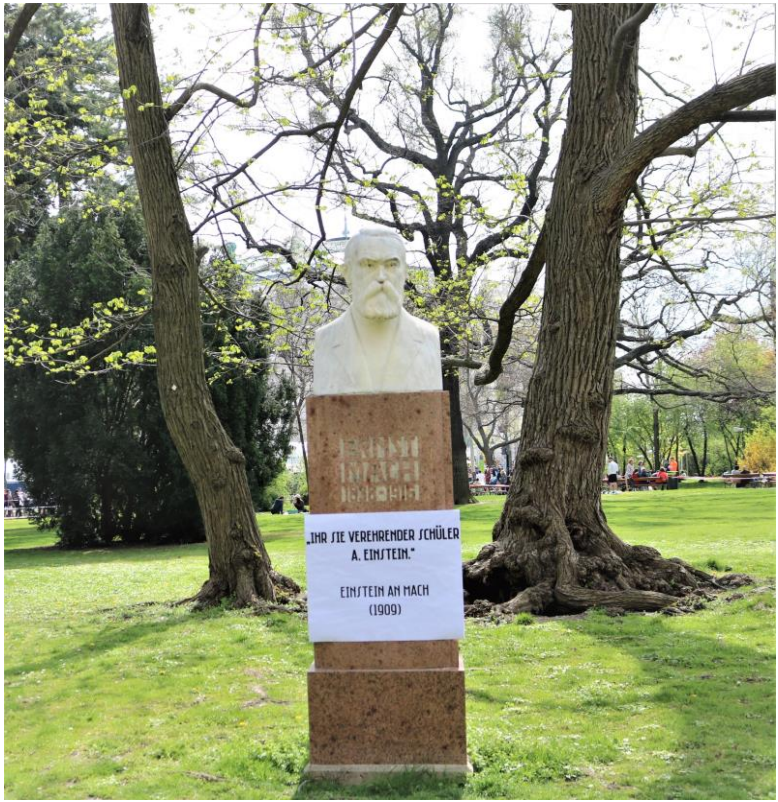

Figure 1

On the same day, the physicist Felix Ehrenhaft – it is noteworthy that he was the only one who nominated Einstein for the Nobel Prize in 1916 – strikingly remarked that Mach´s bust is "alone and isolated" and "not in the courtyard (Arkadenhof) of the University of Vienna, where the busts of many of his famous contemporaries" are (Ehrenhaft 1926, my translation).[xi] The reason for this was probably Mach's preface in his book on optics (*The Principles of Physical Optics*), which appeared posthumously in 1921.

In the foreword he explains that he is "given the role of the forerunner of the theory of relativity." He then goes on saying that he must reject that role and compares the theory of relativity with that of a dogmatic church (Mach 1921, S. VIII). (He also refers to a second volume, where he will explain his criticism. Mach 1921, S. IX) He describes himself as an "unprejudiced walker with his own thoughts" (Mach 1921, S. VIII). Given that foreword, the bust in the park makes sense.

This image created in by the forward has been used in Mach research.[xii] However, in 1987, through the work *Mach I, Mach II, Einstein und die Relativitätstheorie. Eine Fälschung und ihre Folgen.* of Gereon Wolters (1987), it became apparent that the foreword was a forgery[xiii] (and a second volume never existed). In this respect, the monument remains the same, but our image in research has changed since then. This forgery has deceived some in the past- including Einstein. Let us read how he reacted to the forgery.

Arnold Sommerfeld draws Einstein´s attention to the preface in a letter (4 July 1921):

"It is very amusing that Mach takes an *adverse* position on rel. th. in his very fine posthumous work on optics (comp. introduction)[xiv] while, according to the honorable Weyland, you plagiarized him anyway." (CPAE 13; 168; p. 122)

It is not surprising that Sommerfeld mentions Paul Weyland in his letter. After Einstein´s prediction of light diffraction was experimentally confirmed in 1919, making him famous, some opponents of relativity sought attention. It was well known that Weyland was not only antisemitic but also the chairman of the "*Arbeitsgemeinschaft deutscher Naturforscher zur Erhaltung reiner Wissenschaft e.V.*" (Grundmann 2004, p. 98). "This association has gone down in history as the 'Anti-Einstein League'." (Grundmann 2004, p. 98)[xv], which tried to defame the theory of relativity and Einstein. Einstein and his friends (Max von Laue, Heinrich Rubens, Walther Nernst) of course objected. Broadly described, this was the situation in which *The Principles of Physical Optics* by Mach was published posthumously with a fake preface.

Mach, however, was an opponent of antisemitism and considered as a forerunner of the theory of relativity during his lifetime. That is why his posthumously alleged rejection came as a surprise. How much this falsification has burdened the Mach research in a negative way, has already been thoroughly examined by Wolters (1987).

In the course of my paper, I will try to show that this forgery (as well as Planck´s polemic criticism) ironically also has a positive side. Einstein could of course have answered Sommerfeld that his theory had little in common with Mach´s thoughts – thus he could have dismissed both Weyland´s critique of plagiarism and also liberated himself from the shadow of Ernst Mach. Let´s take a look at Einstein´s letter to Sommerfeld (13 July 1921), which in my opinion reflects Einstein´s real grandeur:

"It surprises me that Mach did not favor the theory of relativity. For this chain of reasoning lies entirely along his line of thought. I am curious how he arrives at his disavowal." (CPAE 13; 175; p. 125)

Einstein was surprised because out of the Mach-Planck debate, a Mach-Einstein coalition had formed and Einstein understood himself as a defender of Mach. (Planck and Einstein were very good friends – only in terms of Mach and the general theory of relativity, they were divided.)[xvi] But what exactly does Einstein mean with "For this chain of reasoning lies entirely along his line of thought"? It can be presupposed that Einstein knew Mach´s works and views very well (see Wolters 2019, §28.4). He believed that the theory of relativity, in his own estimation, was consistent with Mach's "line of thought". The question remains whether Einstein has ever expressed his thoughts on this topic directly or indirectly.

Mach´s book with the forged preface appeared, as already mentioned, in 1921. The Nobel Prize for Physics was not awarded in this year. Einstein received the Noble Prize of 1921 retrospectively in 1922 "for his services to Theoretical Physics, especially for his discovery of the law of photoelectric effect" (The Nobel Prize in Physics 1921 [online]). (Whether the fake preface played a role in this topic is currently open to question, by myself at least.) He could not personally receive the Nobel Prize because he was on a trip to Japan at that time. At the request of the philosopher Kitaro Nishida Einstein spontaneously gave a speech "How I created the theory of relativity" (on December 14, 1922 at Kyoto University). Jun Ishiwara, who studied among others at Sommerfeld and Einstein from 1912-1914, translated Einstein´s German speech for the audience at that time into Japanese and published his notes later (in Japanese).

**3. Einstein´s speech**

"Old EINSTEIN at the blackboard stands

His sermon flowing from mouth and hands

While ISHIWARA with sharp, quick look

Writes it all down in his little book. "

(Einstein´s poetic lines about his speech.)[xvii]

Unfortunately, Einstein did not document or publish his speech. Due to the background I described above (§2), it is not surprising. (Finally, a second part *The Principles of Physical Optics* was announced, where Mach allegedly wanted to criticize, among other things, the theory of relativity.) Einstein took advantage of the opportunity of expressing his thoughts on a distant continent.) When translating the Japanese text, there is some room for different interpretations in some places, but those are not so significant for this paper. I will use the translation of the *The Collected Papers of Albert Einstein* for my investigation. In doing so, I will number the lines – with [1], [2] ... etc. – and add some comments, so that my analysis is easy to follow.

"[1] It is by no means easy to give an account of how I arrived at the theory of relativity. […] [2] I will only briefly summarize the key points of the main strand in the development of my thinking.

[3] The first time I entertained the idea of the principle of relativity was some seventeen years ago. [4] From where it came, I cannot exactly tell. [5] I am certain, however, that it had to do with problems related to the optics of moving bodies. [6] Light travels through the ocean of the ether, and so does the Earth. [7] From the Earth's perspective, the ether is flowing against the Earth. [8] And yet I could never find proof of the ether's flow in any of the physics publications. [9] This made me want to find any way possible to prove the ether's flow against the Earth, due to the Earth's motion. [10] When I began pondering this problem, I did not doubt at all the existence of the ether or the motion of the Earth. [11] Thus I predicted that if light from some source were appropriately reflected off a mirror, it should have a different energy depending on whether it moves in the direction of the Earth's movement, or in the opposite direction. [12] Using two thermoelectric piles, I tried to verify this by measuring the difference in the amount of heat generated in each.[2] [13] This idea was the same as in Michelson's experiment, but my understanding of his experiment was not yet clear at the time.[3]

[14] I was familiar with the strange results of Michelson's experiment while I was still a student pondering these problems, and instinctively realized that, if we accepted his result as a fact, it would be wrong to think of the motion of the Earth with respect to the ether.[4] [15] This insight actually provided the first route that led me to what we now call the principle of special relativity. [16] I have since come to believe that, although the Earth revolves around the Sun, its motion cannot be ascertained through experiments using light.

[17] It was just around that time that I had a chance to read Lorentz's monograph of 1895. [18] Lorentz discussed and managed to completely solve electrodynamics to first order approximation, i.e., neglecting quantities of the second order and higher of the ratio of the velocity of a moving body to the velocity of light. [19] I also started to work on the problem of Fizeau's experiment and tried to account for it on the assumption that the equations for the electron established by Lorentz also hold when the coordinate system of the vacuum is replaced by that of a moving body. [20] At any rate, I believed at the time that the equations of Maxwell-Lorentz electrodynamics were secure and represented the true state of affairs. [21] The circumstance, moreover, that these equations also hold in a moving coordinate system gives us a proposition called the constancy of the velocity of light. [22] This constancy of the velocity

---

[2] Whether Einstein conducted the whole experiment or not due to his limited resources is difficult to determine (see Abiko 2000, pp. 8 ff., 13). As I have already mentioned, I follow the translation of Haubold and Yasui 1986. It becomes clear from their translation (see Haubold and Yasui 1986, p. 273) that Einstein only had the idea of creating such an experiment. For this reason, I discuss his elaborations as a preliminary stage of the experiment or in other words as a thought experiment.

[3] See CPAE 13; 399; pp. 639 f., fn. 4 for different interpretations of the translations.

[4] See CPAE 13; 399; pp. 640, fn. 5 for different interpretations of the translations.

of light, however, is incompatible with the law of the addition of velocities known from mechanics.

[23] Why do these two things contradict one another? [24] I felt that I had come upon an extraordinary difficulty here. [25] I spent almost a year fruitlessly thinking about it, expecting that I would have to modify Lorentz's ideas somehow. [26] And I could not but think that this was a riddle that was not going to be solved easily.

[27] By chance, a friend of mine living in Bern (Switzerland) helped me. [28] It was a beautiful day. [29] I visited him and I said to him something like: [30] 'I am struggling with a problem these days that I cannot solve no matter what I try. [31] Today I bring this battle of mine to you.' [32] I had various discussions with him. [33] Through them it suddenly dawned on me. [34] The very next day I visited him again and told him without further ado: [35] 'Thank you. [36] I have already solved my problem completely.'[5]

[37] My solution actually had to do with the concept of time. [38] The point is that time cannot be defined absolutely, but that there is an inseparable connection between time and signal velocity. [39] Using this idea, I could now for the first time completely resolve the extraordinary difficulty I had had before.

[40] After I had this idea, the special theory of relativity was completed in five weeks. [41] I had no doubt that the theory was also very natural from a philosophical point of view. [42] I also realized that it fitted nicely with Mach´s viewpoint. [43] Although the special theory was, of course, not directly connected with Mach's viewpoint, as were the problems later resolved by the general theory of relativity, one can say that there was an indirect connection with Mach´s[6] analysis of various scientific concepts.[7]

---

[5] Hisaki Hashi offers the following additional alternative for the translation of [35]-[36]. (I will try to reproduce her suggestion, as far as I understood.) Instead of the word "solved" one can use a word-for-word translation as follows: "(…) interpreted." (This comes to pass, because of "kaishaku suru" 解釈する. kaishaku 解釈 means "interpretation" as a noun. If one were to use it as a verb, then: "kaishaku suru" 解 釈する. Because of that, the word-for-word translation doesn't sound good in German (or in English) – but the translator may stylize this, in following direction: "Danke. Durch eine bestimmte Interpretation habe ich mein Problem bereits im Ganzen gelöst." Or in English: "Thank you. By a certain interpretation, I already have solved my problem completely."

[6] Einstein profits highly from the thoughts of Mach and Hume. (Several parallels can be found between the thoughts of Hume and Mach.) The fact that Einstein only mentions Mach at that point is mainly connected to the background of the speech.

[7] Haubold and Yasui translate [42]-[43] as follows:

[44] Thus the special theory of relativity was born." (CPAE 13; 399; pp. 636 ff.)

### 3.1. The analysis of Einstein´s speech

"There is no doubt,

that the thought experiment initiates the greatest transformations in our thinking,

and that it establishes the most important avenues of research."

(Mach 1897, p. 2, my translation)[xviii]

The speech of Einstein has already been examined by some philosophers of science but the focus was on the question of whether and how Einstein was influenced by the Michelson experiment. However, as with both the background and the content of the speech itself, it is

---

"Sie stimmt auch mit der Machschen Theorie überein. Solche Probleme jedoch, die später durch die Allgemeine Relativitätstheorie gelöst wurden, standen nicht direkt mit MACHS Lehre im Zusammenhang. Er hatte aber viele wissenschaftliche Begriffe analysiert und verdeutlicht, unter denen es einige gab, welche man als indirekt mit der Relativitätstheorie zusammenhängend betrachten dürfte." (Haubold and Yasui 1986, p. 275)

An English translation would be:
"It also fit in with Mach´s theory. However, such problems, which were later solved by general theory of relativity, were not directly connected with MACH´s viewpoint. But he had analyzed and clarified many scientific terms, among which there were some that might be connected with the theory of relativity."
This translation also makes more sense, according to my analysis. Mach has already emphasized during his lifetime that his considerations are in accordance with special relativity (see Wolters 2019). And Einstein knew that (see Blackmore and Hentschel 1985, pp. 58-59). In addition, Einstein assumed that posthumously Mach had criticized the special relativity in the forged preface (see Speziali 1972, p. 391), which also explains why Einstein wondered about Mach´s preface. That´s why Einstein wants to point out in his speech that Mach´s theory was in accordance with the theory of special relativity. That is why I am trying in this work to connect Mach´s scientific work with the development of the theory of special relativity.

easy to see that it can make a significant contribution to the Mach-Einstein research in particular, what the following analysis will show.

First and foremost, the larger structure reveals itself in the entire speech: it consists of two parts. In the first part [1] - [44] he explains how he arrived at the special theory of relativity and in the second part the general theory of relativity. And both parts show the same structure[xix]:

1.) Phase I: The key idea [3] - [15]

   a) He gives a time when the idea for his theory emerged.[xx] [3]

   b) He indicates which field of physics it is concerned with.[xxi] [5]

   c) He began to question the essential aspects by means of a thought experiment. [11] - [12]

2.) Phase II: The further development of existing concepts [17] - [40]

   a) He examines existing physical concepts. [17] - [20]

   b) He recognizes a contradiction. [22] - [26]

   c) He resolves the contradiction. [37] - [38]

3.) Phase III: Philosophical Authority [41] - [43]

   a) He explains that his theory is in accordance with the philosophy / epistemology of Mach. [41] - [43]

Since part II follows part I directly, one can interpret these phases cyclically. In a sense, Einstein offers, according to my personal assessment, an insight into how the circle of his thoughts closes again. It is not surprising how each part (I and II) ends with phase III. Einstein´s intention becomes clear: the theory of relativity is compatible with Mach´s ideas. This is understandable in the context of the historical and social background. If that is the intention of the speech, the circle thought which closes with Mach could also contain other elements that relate to Mach´s contributions. Maybe the circle even starts with an idea of him. So if one can connect phase I with Mach, then one should be able to reconstruct Einstein´s thoughts. Most importantly one should identify the first ideas that relate to Mach. Since Mach is regarded as a pioneer in the theory of thought experiments (see Kühne 1997, p. 3; Sorensen 1992, pp. 4, 51), it seems likely that the thought experiments are the right place to start with our investigation (phase I c)). In order to do that, however, we must understand how the thought experiments of Einstein are connected with those of Mach to understand all the phases.[xxii]

### 3.1.1. Phase I

On the basis of Einstein's correspondence, it is easy to show that Einstein (as early as 1899), among other things, studied the works of Mach eagerly.[xxiii] In his speech he says that he examined the optics of moving bodies [6]. He then explains his thought experiment with mirrors [11] - [12] and connects it with the experiment von Michelson [13] - [15]. Now, of course, the question arises: Has Mach ever worked on optics?

Most of Ernst Mach´s published work belongs to the field of basic scientific research, including his experiment with which he confirmed the Doppler effect in 1860. The Doppler effect has become an indispensable part of the scientific world and even arrived in pop culture. But as it is the case with all new concepts, the thoughts and arguments of Christian Doppler were not fully matured at his lifetime and had several critics. It will become Mach´s task, after the death of Doppler, to finalize his thoughts both theoretically and experimentally and thus to end the discussion between the physicist Doppler and the mathematician Joseph Petzval – in favor of Doppler. The scope of the paper does not allow me to discuss the work of Doppler and Mach on this subject in great detail.[xxiv] Nevertheless, I will briefly focus on Mach´s publication *Beiträge zur Doppler´schen Theorie der Ton- und Farbänderungen durch Bewegung. Gesammelte Abhandlungen*[xxv] (1873) (*Contributions to Doppler's theory of sound and color changes through motion. Collected essays*)[xxvi] because, like Susan G. Sterrett in her article "Sounds Like Light: Einstein's Special Theory of Relativity and Mach's Work in Acoustics and Aerodynamics" (1998), I think it is very likely that Einstein was aware and studied this work of Mach, although he never explicitly mentioned reading it.

The reason for my assumption is, as I will discuss in more detail below, that the Doppler effect was very important for the genesis of special relativity. One had to analyze exactly how the analogy between light and sound waves was to be understood. And if one wanted to understand this analogy in physics, Mach was as an expert and potential consultant. An example is the correspondence between Ernst Mach and Otto Neurath (1914-1916). Neurath was thankful for the "friendly lines" of Mach (Haller 1980, p. 3; Blackmore and Hentschel 1985, p. 150). Subsequently, Neurath wrote that he will send Mach his lecture "Zur Klassifikation von Hypothesensystemen (Mit besonderer Berücksichtigung der Optik)" ("On the Classification of Hypothesis Systems (with special consideration of optics).") (The lecture was given by Neutrath on March 2, 1914 (see Haller and Rutte 1981, p. 85). Then Neurath asked Mach, whether he could tell him about "The light is something like sound," "The sound something like the light." (Haller 1980, p. 4). For three reasons, I believe that Neurath, in his request had thought of the development of the special relativity:

(i) Neurath analyzed in his lecture "On the Classification of Hypothesis Systems (with Special Consideration of Optics)" (Haller and Rutte, p. 88, my translation) the history of optics from the beginning of the 17[th] century until the beginning of the 19[th] century (Haller and Rutte 1981, p. 88). He referred to the analogy between sound and light (Haller and Rutte 1981, pp. 91, 96 ff.). (Mach is mentioned several times in this work (see Haller and Rutte 1981, pp. 86, 101)). Neurath wrote

(ii) another letter to Mach: "In this lecture I have extended the philosophical results more broadly, and I am currently pursuing some of the problems indicated here." ( Haller 1980, p. 5, my translation) It is possible that he therefore intended to study the history from the entire 19$^{th}$ century to the beginning of the 20$^{th}$ century.

(ii) In the same letter, in which he asked Mach for his view on the analogy between sound and light, Neurath wrote, that he is following the implementation of Mach's thoughts into general relativity with the utmost interest (see Haller 1980, p. 4). (I will expand on this in my work "On thought experiment: Mach and Einstein (Part II)".)

(iii) Neurath also seemed to be interested in Mach´s views even after their correspondence. Therefore, he was also in possession of *The principles of physical optics* (1921) with the forged preface. On the first page of the book he noted: "M. stresses the point, that he is not a forerunner of the theory of relativity, and against the atomistic belief of today p. VIII". (fig. 2)[xxvii]

The fact that he underlined this note further shows for me that he did not expect such a view of Mach.

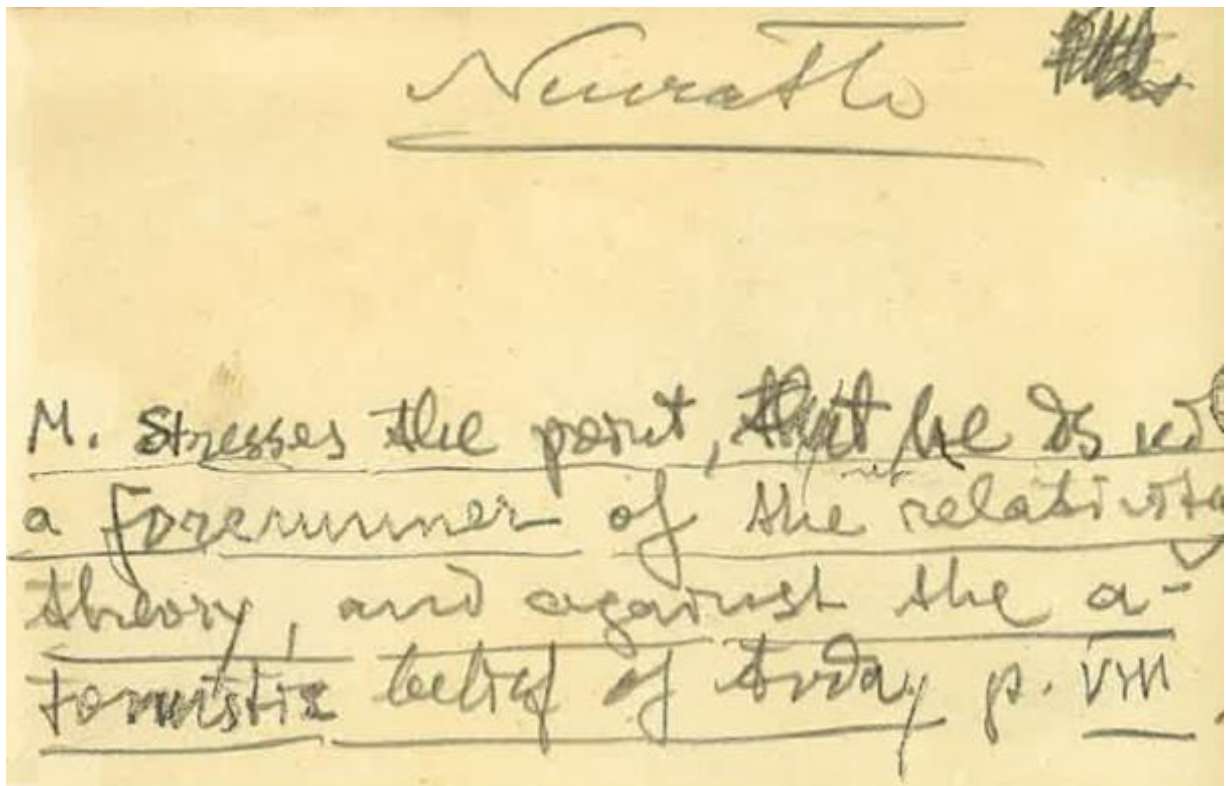

*Figure 2*

Based on these arguments (i, ii, iii), I believe that Neurath had probably already recognized that there is a connection between the work of Mach and special relativity. But Neurath might not have been the only one, who recognized this. Max von Laue recognized the close connection between the Doppler effect and special relativity very early. As Einstein writes in "On the Principle of Relativity and the Inferences Derived from It" (1907):

"In the first section, in which only the kinematic foundations of the theory are applied, I also discuss some optical problems (Doppler´s principle, aberration, dragging of light by moving bodies); I was made aware of the possibility of such a mode of treatment by an oral communication and a paper by Mr. M. Laue (*Ann. d. Phys. 23* (1907): 989), as well as a paper (though in need of correction) by Mr. J. Laub (*Ann. d. Phys. 32* (1907))."

(CPAE 2; 47; p. 254)

It is interesting that exactly this article is mentioned in Einstein´s letter to Mach (1909) (CPAE 5; 175; p. 130). In any case, it is not surprising in this context, what we read in the article "On Mach´s *The Principles of Physical Optics*" (1921) by Max von Laue:

"In this book, a man of undeniable significance speaks perhaps for the last time to his contemporaries ('perhaps' because a continuation is planned which hopefully will also come out). One will therefore certainly take it in hand with utmost respect. […] Though the intent is to place it besides the historical-critical meachanism by the same author, it has nothing of the originality which that the 30 year old work had overall, and nothing of the scientific insight, that very frequently shows. […]

But the preface is nothing short of 'sensational'. The author provisionally – the exact details will follow in the second part – takes a position with respect to the theory of relativity, and indeed, a rough, unfriendly one. [He sees that people] are gradually beginning to think of him as a 'forerunner to the theory of relativity', and he believes that he must reject this [identification] with the same decisiveness with which he has denied 'the atomistic dogma of the current school or church' for his own person. For the author, even relativity theory seems 'dogmatic', and wherever he spots that word, he senses (as can be noticed in more than one place in the book which is before us) an utterly unhistorical, uncritical, horror.

If one were to characterize the situation, it cannot be correct to push skepticism so far that one denies one's own deepest thoughts the moment a person greater than oneself develops it into something positive. And yet the book has rendered a service to the theory of relativity: He never speaks of a palpable 'Aether' and shows how well all of wave optics can be represented without it." (von Laue 1921, p. 65-66)

Von Laue emphasized that Mach did not write about the ether and indirectly suggested that Mach and Einstein´s considerations are somehow related. But how are they related?

Dieter B. Herrmann wrote in his work "Erkenntnis und Irrtum: CHRISTION DOPPLER" ("Knowledge and Error: Christian Doppler") (1964):

"In this context, the opinion of the great Viennese physicist ERNST MACH is interesting. He, who also contributed considerably to the history of the theory of relativity, came closest of all physicists to the real meaning of the DOPPLER principle."

(Herrmann 2004, p. 50, my translation)[xxviii]

Unfortunately, Herrmann did not elaborate on how Mach had "contributed considerably to the history of the theory of relativity" – But if you look at Neurath´s letter together with Herrmann´s article, then you have a good hint: Mach´s work on the Doppler effect. At this point we want to list some important achievements of Mach in this area (Doppler effect):

1. Mach is the first, who introduces a replicable experiment (Machsche Pfeife)[xxix] for the acoustic Doppler principle.
2. He sees the formula of Doppler just as an approximation law (Mach 1873, p. 9).
3. He discusses the Doppler effect based on the perspectives of various observers (Mach 1873, pp. 6 f., 13).
4. His work, strictly speaking, like that of Christian Doppler, belongs not only to the field of physics, but also to astrophysics. In astrophysics he does not only assume an optical Doppler effect due to the relative stellar motion, but also discusses the possibility of determining the radial velocity of the stars by measuring the line shifts in star spectra (Mach 1873, pp. 1, 17, 33).
5. Mach tries not to associate the ether with the Doppler effect.

(In this work, in the form of P1, P2, P3, P4 and P5, I will refer of these points so that my analysis become understandable).

Let´s start with the third point (P3). Discussing physical phenomena from the perspective of observers is characteristic of Mach´s thinking. Perhaps it was precisely the Doppler effect that shaped this type of physical reasoning in him. Mach can solve the dispute about the Doppler effect between Petzval and Doppler on the basis of various observers (see Thiele 1971, pp. 398 ff.). Likewise, in Einstein´s thought experiments, the different observational points of view become characteristic of his way of arguing. (It is already well known that Einstein contrasted different observers in his thought experiments. His thought experiment (with the elevator), which can also be found in Einstein´s speech, was also inspired by Mach. I will go into more detail in my work "On thought experiment: Mach and Einstein (Part II)"). In this text I would like to reduce my focus only to special relativity. If we look at Mach´s work on the Doppler effect (1860), then the first experiment he gives are the train experiments of Buys Ballot and M. Scott Russel (Mach 1860, p. 11). Namely, the sound of an arriving train is higher, and as it moves away from us, it goes lower. So that the sound is relative – that is determined by the observation of the observer or measurement.

Important is the tone that you can measure in the place where you are. Mach has even invited scientists to Prague so that they can hear the incoming and outgoing train at the station (see Motz 1988, p. 87). But not only Christian Doppler and Ernst Mach were active in Prague. Einstein´s career as a university professor began in Prague (1911-1912). At this time he published his article "Die Relativitäts-Theorie" ("The Theory of Relativity") (1911), where he explains the relativity principle with a thought experiment right at the beginning – the thought experiment deals with two physicists: one in the open field and the second one goes by train (CPAE 3; 17; p. 340). On the next page he will immediately compare light with sound waves (CPAE 3; 17; p. 341). But that is not the only parallel. Einstein begins his work

"On the Electrodynamics of Moving Bodies" with the critique of simultaneity (Einstein 1905, §1). (In this work, I will discuss below, that for the critique of simultaneity or absolute time,

Einstein was probably inspired by Mach.) Interestingly, Einstein´s first example in this chapter is a train (Einstein 1905, §1, p. 893). Einstein emphasizes that the definition of time in one place is sufficient for what we observe or measure on a clock (Einstein 1905, §1, p. 893):

"Such a definition is indeed sufficient if time has to be defined exclusively for the place at which the clock is located; but the definition becomes insufficent as soon as series of events occurring at different locations have to be linked temporally, or – what amounts to the same – events occurring at places remote from the clock have to be evaluated temporally."

(CPAE 2; 23; p. 142)

One solution, according to Einstein (Einstein 1905, § 1, p. 893), would be that an observer could set up his watch by means of light signals from another observer – but that is associated with a problem, as he emphasizes. Let´s take a closer look at the example. There is an observer A and an observer B (Einstein 1905, § 1, p. 894), who move at a relative speed to each other. If we assume that the speed of light is a constant, then it always needs the same time for the distance between A to B and B to A (Einstein 1905, § 1, p. 894). Suppose an observer (A) had a clock with the frequency (f) of light in his system synchronized with his clock.

If he now sends the light to another observer (B) who has a different relative speed, then the observer (B) receives a light with a different frequency (f´) due to the Doppler effect. Since the frequency is the inverse of the period, it would be understood that the time is transforming. The relativistic Doppler effect thus becomes understandable if one sees the origin of the phenomenon in space and time. In the next chapter Einstein writes "§2. *On the relativity of lengths and times*" (Einstein 1905, §2, p. 895). It was important for Einstein to recognize that the relativistic Doppler effect depended only on the frequency of the emitted signal and the relative velocity between A and B. But for that he should ignore the ether, which we will discuss below.

Yet another point should be mentioned in this context: It is interesting how both Mach and Einstein always address the different perspectives of the "observers". That Mach uses the term "observer" ("Beobachter") (for example Mach 1860, p. 7 f.) in the Doppler effect is understandable, since the Doppler effect directly affects our senses – we can either hear or see the Doppler effect. It is not self-evident that Einstein speaks in his investigation of the time of "observer" (Einstein 1905, § 1, p. 893 ff.), although it would be sufficient to mention coordinate systems. Below, we will see from Besso´s letter to Einstein that Einstein may have been inspired by Mach for his analysis of simultaneity and the relativity of lengths and times. In this case, Einstein´s reception of Mach would have left traces in his text. And we can possibly recognize such: In the introduction, he does not use the term ("observer"). In §1 (topic: definition of simultaneity) he uses the term "observer" three times. In §2, which deals with the relativity of lengths and times based on the relativity principle, he will use it four times. (As I will explain below, it will be precisely the principle of relativity that can be associated with Mach.) In chapters §3, §4, §5, §6 he does not use the term a single time, as expected! The next time the term "observer" reappears – as we may well expect – is §7 (Topic: Theory of Doppler's Principle and Aberration). There Einstein uses the term "observer" six times and thus the most in his work! (He will use it only once in §8 (Einstein 1905, p. 914), and in chapters §9 and § 10 he does not use the term a single time.) This simple conceptual analysis shows just how much the chapters §1, §2 and §7 in Einstein´s mind probably belonged together. This can easily be explained if we attribute a special place to the Doppler effect in the emergence of special relativity. In addition, the assumption that Einstein knew Mach´s work on the Doppler effect well, is very obvious.

It is also interesting that we can find an experiment with mirrors, which Mach suggests measuring the Doppler effect of light (fig. 3), also in Einstein´s work. A description of the experiment can be found in Mach´s work (Mach 1873, p. 32). Because of Mach´s restricted material resources at that time, he just suggests this experiment. It provides a preliminary stage of experimentation, which according to Mach´s definition has to be regarded as a thought experiment[xxx] (Mach 2011, pp. 197 f., 211 f. – see also Marco Buzzoni (2019, p. 660 f.), who had a similar view)[xxxi] – and therefore, it is important to discuss it the following paragraphs.

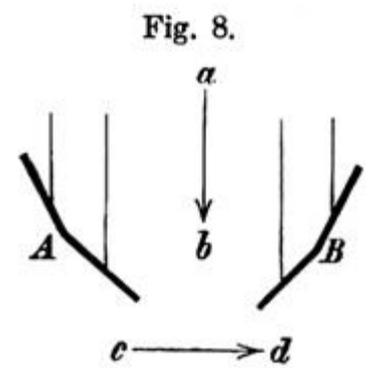

Figure 3

Einstein probably has been inspired by this thought experiment of Mach. He discusses this thought experiment (slightly modified) at least twice, without mentioning Mach.[xxxii] First, it encounters us in his book *Über die spezielle und die allgemeine Relativitätstheorie* (1916)[xxxiii] (*Relativity: The Special and the General Theory*). Second he mentions it in his speech. (In both mentions relating to the Michelson experiment.)

Einstein writes about this thought experiment (Einstein 2009, p. 35) after a discussion of how the Doppler effect can be measured from spectral lines (Einstein 2009, p. 33). As I have mentioned above, Mach already suggests how to measure the Doppler effect with spectral lines (see P4). One page before Mach describes the thought experiment about the Doppler effect, he elaborates on the Machsche Pfeife (Mach´s pipe – see P1) (Mach 1873, p. 31). Einstein argues analogously to Mach about the sound of an organ pipe (Einstein 2009, p. 10) but again, does not mention him at this point. (See also Sterrett 1998, pp. 20 ff., where the connections are explained.)

But let´s get back to Mach's thought experiment. Mach draws attention to an experiment by Fizeau, which inspired his thought experiment (Mach 1873, p. 31 f.). Mach´s aim was, as he explains, to demonstrate with this experiment the Doppler effect of light. The light shines from *a* to *b* (see fig. 3), from the Sun to the Earth. Simultaneously, the Earth moves from *c* to *d* (see fig. 3). The incoming light is reflected by the interference mirrors A and B. Due to the Doppler effect, a difference in the wavelength of the reflected light rays can be found and measured (Mach 1873, p. 32).

The thought experiment thus has the purpose to measure the *relative movement* of the Earth to the Sun with the help of the Doppler effect – see the parallels in Einstein 2009, p. 36 and in Einstein´s speech [16]. Einstein´s comment on the relative motion between the Earth and the Sun is easy to understand, if one considers it an answer to Mach´s considerations. Even more remarkable, however, is that Mach´s argument for the Doppler effect of 1860 does not deal with ether[xxxiv], or omits it[xxxv] altogether (see P5). Moreover, he uses the particle model[xxxvi] in addition to the mathematical wave model. The fact that Mach brings particles of light into the discussion is an ingenious trick that may directly be considered a preliminary to Einstein's quantum light hypothesis.

Mach´s investigations on the Doppler effect in acoustics as well as in optics regard only the relative movements between the observer (= receiver) and the source. If the relative velocity of both the source and the observer is much slower to the medium than the propagation of the waves (speed of sound or speed of light), it can be neglected because they contribute little to the frequency change (see P2).

The relative speed of the Earth is many orders of magnitude smaller than the speed of light. In this respect, the work of Mach can be read in such a way that all systems that move at a relative speed are equal in terms of physical laws.

But if one wants to include the propagation medium / ether in their investigations, his concept and the experiment would need to be developed further. The question arises, why Mach did not make these considerations himself? The answer to this question is the same as to why he did not try to reconcile mechanics with the Maxwell equations (to further study the Doppler effect and aberration) ... namely: Mach´s thought experiment is from the year 1862.

James C. Maxwell's important work *A dynamical theory of the electromagnetic field*, in which his equations can be found, appears in 1865. (He has already published articles regarding this topic between 1861-1864.) It will then take several more decades for these 20 original equations to form what we now know as Maxwell´s equations. Thus, it is not surprising that Ernst Mach, an experimental physicist, did not pursue this question in his work, where he focuses primarily on the experimental proof of the Doppler effect. (In the following years, he will devote himself to other tasks in various fields.) And as John Norton (Norton 2004, p. 3) already emphasizes, electrodynamics was a prerequisite for the special theory of relativity. Einstein seems to point this out in his obituary of Mach:

"It is not improbable that Mach would have hit on relativity theory when in his time – when he was in fresh and youthful spirit-physicists would have been stirred by the question of the meaning of the constancy of the speed of light. In the absence of this stimulation, which flows from Maxwell-Lorentzian electrodynamics, even Mach´s critical urge did not suffice to raise a feeling for the need of a definition of simultaneity for spatially distant events."
(CPAE 6; 29; p. 144)

Einstein himself clearly identifies what was missing in Mach´s work for historical reasons: A theoretical examination of electrodynamics, which would have led him first to the constancy of the speed of light and then to a treatment of simultaneity. It is precisely these stages that Einstein discusses next in his speech, which we will discuss in phase II and phase III.

If we now imagine the young Einstein, who thinks about this thought experiment of Mach, the situation can be easily understood. In his thought experiment, Mach simply presupposes the relative motion of optics because he wants to study the relative motion of the Earth to the Sun. But if we look at this experiment from the perspective and the problems of electrodynamics, the following questions arises: What role does the ether play exactly? What about the "absolute" movement of the Earth in relation to the ether?

To answer these questions, one cannot take the Sun (or another star) as the light source because this could always only be interpreted as a relative movement of the Earth to a star. Treder explains: "An absolute determination of the movement of the Earth can only work with the light of earthly sources, [...]" (Treder 1982, p. 93, my translation). For this reason, it is understandable why Einstein remodels the thought experiment. In general, Einstein probably had several ideas of the experiment planned (see Stachel 1982, p. 50 and Pais 1982, p. 133).

In Einstein´s thought experiment, the mirrors are set differently and are provided with thermocouples. The light source in his experiment is not the Sun. Furthermore, the light is split in opposite directions. Einstein´s question is whether someone can measure a velocity of the Earth in relation to the ether with the help of the Doppler effect [8] - [12]. At the latest after the result of the Michelson experiment it must have been clear to him that this is not possible and thus the principle of relativity applies. For this reason, Einstein discusses the Michelson experiment in his speech in connection with the principle of relativity [16] and not in connection with the constancy of the speed of light. It is interesting that John Stachel comes to the same conclusion in his study of Einstein´s Kyoto speech but he only sees it as a "justification" (see Stachel 1982, pp. 48 f.).

The work of Abiko 2000 and Dongen 2009, as well as this study, have shown that the Michelson experiment was relevant to Einstein in the development of the special theory of relativity (for a different view see Stachel 1982 and Holton 1988). But whether the Michelson experiment should be considered as a crucial experiment (Treder 1982) or not (Holton 1988, pp. 279 ff.) depends on the theoretical framework researchers are referring to when they reconstruct the history of science. I would therefore argue for a pluralistic history of science. H.-J. Treder´s paper, which he published in 1982, is very worth reading because he sees the Michelson experiment as an experimentum crucis, but nevertheless does not necessarily consider the ether model as generally eliminated and further discusses it in the field of quantum field theory. (By the way, Einstein refers to Fizeau´s experiment as "experimentum crucis" (CPAE 3; 2; p. 118), and was later happy when physicists devoted themselves to this experiment (see CPAE 8; 109; p. 161).

In none of the works above Mach is mentioned or discussed in the context of special relativity. This is not surprising because they all focus on Michelson's experiment. But as I clarified above, an interesting historical background is hidden in Einstein's thought experiment. That Einstein probably took and transformed the thought experiment of Mach serves as a direct indication of the area and the topic Einstein started his investigations on the special theory of relativity with: the field of optics, more precisely with the Doppler effect. (In this respect, a certain irony may be apparent in Einstein´s speech: If Mach´s thought experiment on the optics was part of Einstein´s inspiration for the genesis of special relativity, then we can imagine how much the forgery preface in *The Principles of Physical Optics* must have surprised him…)

Since Mach in contrast to Doppler does not investigate the subject of aberration, we can also immediately guess what Einstein will do next: He tries to find out how the Doppler effect and the aberration are related. This leads us to phase II.

### 3.1.2. Phase II

Of course, one could be skeptical of the thesis that the Doppler effect was of great importance for the development of the special theory of relativity. But if we look at the history of science, we can see that Einstein was not the first who followed that line of thought.

For example, the path from the Doppler effect to the special theory of relativity is not very far. This is attested by the work *Ueber das Doppler'sche Princip* (1887) (*On the Doppler*

*Principle*) by Woldemar Voigt. Andreas Ernst and Jong-Ping Hsu explain in their article *First Proposal of the Universal Speed of Light by Voigt in 1887* that this work "contains several original and fundamental ideas of modern physics". Voigt introduced "the idea of the universal speed of light" and "postulated the invariance of a physics law, the wave equation in 'an elastic incompressible medium'" "to derive the Doppler effect". Furthermore, he analyzed "that the Doppler shift of frequency is incompatible with Newtonian absolute time [...] and is in harmony with a 'relative time'"; and he "first derived a type of 4-dimensional space-time transformation" (Ernst and Hsu 2001, pp. 211 f.).

Despite his important contribution, Voigt has fallen into oblivion these days – a short overview of his work is given in the paper "Recognition for Waldemar Voigt" (Doyle 1988) by William T. Doyle. H.A. Lorentz himself pointed to the pioneering role of Voigt – as so he writes in his paper *Elektrogmagnetische Erscheinungen in einem System, das sich mit beliebiger, die des Lichtes nicht erreichender Geschwindigkeit bewegt* (1913) (*Electromagnetic phenomena in a system moving with any velocity smaller than that of light*) in a footnote:

"It is Einstein´s credit to be the first person to have verbalized the principle of relativity as a general, strict and precisely valid law.

I add in the remark that Voigt in 1887 (Göttingen Nachrichten p. 41) in his work 'On the Doppler Principle' had already applied a transformation on the following equation

$$\Delta\psi - \frac{1}{c^2}\frac{\partial^2\psi}{\partial t^2} = 0$$

which is contained in equations (4) and (5) of my work. (Note by H.A. Lorentz, 1912)" (Lorentz 1913, p. 10, fn. 1, my translation)[xxxvii]

Lorentz thus indirectly refers to the importance of the Doppler effect for the special theory of relativity. Unfortunately, Voigt (similar to Lorentz) gives the ether a special role in his investigation in contrast to Mach.

Likewise, Einstein, to whom Voigt among others was not a stranger (CPAE 1; 126; p. 184), will not give up the concept of the ether for some time and will include it in his theory of "electrodynamics of moving bodies" (CPAE 1; 128; p. 187). Obviously, Einstein was not quite ready at that time.

This is also evident in his speech. He deals with the work *Attempt of a theory of the electrical and optical phenomena in moving bodies* by H.A. Lorentz [17], which helped him significantly in his theoretical research. The third chapter "Section III. Investigation of oscillations excited by oscillating ions." (Lorentz 1895, p. 48 ff.) of the book is important because right at the beginning Lorentz discusses "*local time*" ("*Ortszeit*") (Lorentz 1895, pp. 51 f.). Einstein will later reinterpret the concept of local time in order to arrive at a solution (see Haubold und Yasui 1986, p. 275, fn. 8). At the end of this chapter "The law of Doppler" (Lorentz 1895, pp. 56 ff.) is also discussed in this context. But, as we can guess, Einstein also deals with aberration. Lorentz examines, where he refers to the results of Doppler's law, "The aberration of light" in "SECTION V" (Lorentz 1895, pp. 88 f.). (How the Doppler effect is related to aberration, I will not go into more detail here. A good insight, for example, offers Einstein 1905, §7; von Laue 1919, pp. 118 ff.; Sonne and Weiß 2013, pp. 259 ff.) To understand the relativistic aberration one absolutely needs the formula of the relativistic velocity addition. Einstein certainly has been thinking intensively about this connection and tried to interpret it for his theory. We learn more in detail about it from the conversation between Shankland and Einstein:

"When I asked him how he had learned of the Michelson-Morley experiment, he told me that he had become aware of it through the writings of H. A. Lorentz, but *only after 1905* had it come to his attention! 'Otherwise,' he said, 'I would have mentioned it in my paper.' He continued to say the experimental results which had influenced him most were the observation on stellar aberration and Fizeau´s measurements on the speed of light in moving water. 'They were enough,' he said." (Shankland 1963, p. 48)

However, later (1952) Einstein emphasized that he now remembered that he knew the Michelson experiment before 1905, namely through the works of Lorentz (Shankland 1963, p. 55). (Also Fizeau´s experiment, which was so important to Einstein, was repeated by Michelson and Morley, to which Lorentz points in section V (Lorentz 1895, pp. 99 f.).

I think this interview from Shankland gives a good idea of what Einstein meant in his speech when he refers to Lorentz and Fizeau [19] – [21]. Due to the aberration he comes to a velocity addition, but it does not fit with the mechanics – a contradiction can be observed [22] – [23] and he thought for a long time about the problem according to his speech [24] – [26].

What he lacked was not physical knowledge, but an enlightened, critical point of view. To develop such a critical perspective took him until 1905. A 'revolution' still had to take place in his understanding.[xxxviii] (I personally prefer to describe this change of thinking as an 'evolution' or development in his mind rather than a revolution.) Conversations with his friends certainly helped him a lot – it hardly surprises at this point that his best friends, like Friedrich Adler (Feuer 1971, pp. 282 ff.) and Michele Angelo Besso (Feuer 1971, pp. 291 ff.) were deeply convinced Machians. Especially Besso seems to have been very relevant for the special theory of relativity (see [27] – [36]) – Einstein ends his monumental work *On the Electrodynamics of Moving Bodies* (1905) with the words:

"In conclusion, let me note that my friend and colleague M. Besso steadfastly stood by me in my work on the problem here discussed, and that I am indebted to him for many a valuable suggestion.

Bern, June 1905. (Received on 30 June 1905)" (CPAE 2; 23; p. 171)

Besso is the only one whom Einstein mentions in his acknowledgement – but what did he imply by saying "for many a valuable suggestion"?

More information about this we can read in a letter from Besso to Einstein (from 1947): "My dear, best, old friend.

[...]

As far as the history of science is concerned, in my opinion Mach has been at the center of the development of the last 50 or 60 years. Is it true: - that a somewhat older engineer [Besso], about 1897 or 98, points the young, exceptional scientifically interested [Einstein], who was kept in tension by the questions about the tangibility of the ether and the atoms in tension, towards Mach; and: that this advice was given in a development phase of the young physicist, where Mach's trains of thought pointed decisively to the observable - perhaps indirectly, on 'clocks and scales'?

[…]" (Speziali 1972, p. 386, my translation)

Besso´s help to Einstein becomes apparent. Besso is, in a sense, the conscience of Einstein in these and other lines. And indeed, Ernst Mach discusses and criticizes in his book *Die Prinzipien der Wärmelehre* (*Principles of the Theory of Heat*) the following concepts: 'absolute space', 'absolute time' and especially 'simultaneity'. Moreover, he gives a reference to his book *Die Mechanik in ihrer Entwicklung* (*The Science of Mechanics*). A closer examination of this topic can be found in the introduction of *Die Prinzipien der Wärmelehre* by Michael Heidelberger and Wolfgang Reiter (Mach 2016, pp. XXVII ff.).

Interesting in this context is also another well-known thought experiment of Einstein. He begins namely *On the Electrodynamics of Moving Bodies* with the following words:

"It is well known that Maxwell´s electrodynamics – as usually understood at present – when applied to moving bodies, leads to asymmetries that do not seem to attach to the phenomena. Let us recall, for example, the electrodynamic interaction between a magnet and a conductor. The observable phenomenon depends here only on the relative motion of conductor and magnet, while according to the customary conception the two cases, in which, respectively, either the one or the other of the two bodies is the one in motion, are to be strictly differentiated from each other." (CPAE 2; 23; p. 140)

It has already been recognized or suspected (see Norton 2004, p. 44, fn. 10) that Einstein probably took over and analyzed his thought experiment with the magnet and a conductor from the book *Einführung in die Maxwell´sche Theorie der Elektricität* (*Introduction to Maxwell´s Theory of Electricity*) by Föppl. It is important to ask the question: Why did Einstein, in 1905, employ this book and this chapter from Föppl? A possible answer is briefly drawn out here: as we have discussed above, Besso would have most likely referred Einstein to Mach´s critique and its concepts before the completion of the special theory of relativity. (If we trust Einstein´s speech, it may have been in May 1905.)[xxxix] So Einstein had reason to study the works of Mach or the works investigate with Mach´s concepts. It is interesting that Föppl submitted a paper in 1904 (November 5) entitled *Über absolute and relative Bewegung* (*On Absolute and Relative Movement*). (This work should have been published no later than the beginning of 1905.) That this work and its title must have been of interest to Einstein for many reasons, I think, goes without saying. Let´s take a look at how Föppl introduces his work *On Absolute and Relative Movement*:

"The most accurate statements about the physical meaning of the law of inertia and the related concept of absolute motion derive from Mach. According to him, in mechanics, and as in geometry, the assumption of an absolute space and thus an absolute movement in the true sense is inadmissible. Every movement is understandable only as a relative, and what is usually called absolute motion is merely the movement relative to a frame of reference, a so-called inertial system, required by the law of inertia [...]" (Föppl 1904, p. 383, my translation)[xl]

(I will discuss this work by Föppl in "On thought experiment: Mach and Einstein (Part II)".) When Einstein has read Föppl´s work, I think it is understandable that he also takes a deeper look in *Einführung in die Maxwell´sche Theorie der Elektricität* (*Introduction to Maxwell´s Theory of Electricity*). In this book there is a fifth section "Die Elektrodynamik bewegter Leiter" ("The Electrodynamics of Moving Conducturs") (Föppl 1894, p. 307). There, right at the beginning, he discusses "§ 114. Relative und absolute Bewegung im Raume" ("§ 114. Relative and Absolute Movement in Space") (Föppl 1894, p. 307 ff.). And explains his thoughts on the relative movement between the magnet and the conductor.

The attentive reader will certainly not have missed the parallels between these titles "Über absolute and relative Bewegung" and "§ 114. Relative und absolute Bewegung im Raume". If Einstein knew by Föppl's work (1904), then he would also have realized that one had to rethink again the relative movement of the conductor and magnet. And I think that´s what Einstein most likely did. So, once again, a circle of thought closes: Einstein thanks his good friend Besso at the end of his work. As we can see from Besso´s letter to Einstein, Besso generally sees his contribution as pointing to Mach´s considerations. Einstein´s work begins with reflections on magnets and the conductor, which he probably became aware of as a result of his research on Mach. The beginning and the end of Einstein´s work thus enclose the content: physically examining Mach´s considerations and fitting them into the state of research. (Ernst Mach was an experimental physicist and at that time (1905) already very old and ill. A young, gifted theoretical physicist, like Einstein, was needed for the ideas of the relativity of time, space, and inertia at least for inertial systems as a first step, which allowed him to unify mechanics with electrodynamics.)

Let's move on to the final phase ...

### 3.1.3. Phase III

Einstein ends his speech (part I) with the words:

"After I had this idea, the special theory of relativity was completed in five weeks. I had no doubt that the theory was also very natural from a philosophical point of view. It also fitted with Mach's theory. However, such problems, which were later solved by the general theory of relativity, were not directly connected to Mach's viewpoint. But he had analyzed and clarified many scientific terms, among which there were some that might be connected to the theory of relativity.

Thus, the special theory of relativity was born." (A possible Translation.)

Now it could be considered a historical coincidence that the investigations on the Doppler effect led to others approaching the special theory of relativity already before Einstein, that the thought experiment (of Mach) about the Doppler effect is mentioned by Einstein, that Einstein´s friends were Machian and that Einstein studied himself Mach intensively etc.

But one could also conclude, that these correlations are more than just coincidences. It should not be dismissed, that if someone wanted to physically understand light at that time, he or she had to start with the Doppler principle. One had to first ask the question of how this analogy between light and sound was to be understood and whether the analogy between air and ether was justified before moving on to more complex considerations. Finally, it was also necessary to understand (also by performing thought experiments) where these analogies led too far. From this perspective, it is not surprising that H.A. Lorentz published and annotated the works of Christian Doppler (Doppler 2000). (Namely, the scientist to whom the Lorentz transformation of the special theory of relativity is specifically attributed.)

It will also be Lorentz, together with the physicist W.H. Julius (1912), who proposes Mach for the Nobel Prize for physics. In doing so, they draw attention to the contributions of Mach to acoustics and optics, and to the important role Mach has for the philosophy of science. Finally, he points in this context to Mach´s profound influence on modern ideas and on many physicists (Blackmore und Hentschel 1985, pp. 95 f.).

This development can likewise be found in the history of physics. Because what distinguishes the sound waves from light waves is precisely their relationship to the medium in which they spread. Doppler and Mach give the formulas for the Doppler effect, which apply only in the reference system of the transmission medium. For the acoustic Doppler effect, the relative movement to the medium can indeed be skipped in many cases. But not in every case the medium can be omitted: Sound waves make no sense in a vacuum or are even impossible. If, for instance, a (strong) wind is blowing, the relative movement to the air is also important to consider. In this case, you must add or subtract to the velocity of the wave also the wind velocity. The velocity of the wind causes a frequency shift. According to the relativistic Doppler effect, such a transmission medium does not exist for light.[xli] As a result, the (longitudinal) relativistic Doppler effect is completely symmetrical in contrast to the acoustic Doppler effect. Light achieves its constant natural speed in a vacuum, where no sound waves exist. In the vacuum no ether wind blows. This does not allow us to determine whether the (signal) source or the observer is moving. That´s what the Michelson-Morley experiment showed.[xlii] All inertial systems are equivalent.

If one wanted to compare or possibly even connect the mechanics and the electrodynamics[xliii], the Doppler principle is the (inevitable) border area, which one had to examine more closely (see Einstein 1905, §7). Einstein thought in a similar way to Mach and tried to investigate and understand the border areas (see Renn 2004, pp. 50 ff.; Renn 2007, pp. 30 ff.). Einstein writes on the first page of his work (*On the Electrodynamics of Moving Bodies*):

"Examples of a similar kind, and the failure of attempts to detect a motion of the earth relative to the 'light medium', lead to the conjecture that not only in mechanics, but in electrodynamics as well, the phenomena do not have any properties corresponding to the concept of absolute rest, but that in all coordinate systems in which the mechanical equations are valid, also the same electrodynamic and optical laws are valid, as has already been shown for quantities of the first order." (CPAE 2; 23; p. 140)

From today´s point of view, the path is not so difficult: One takes the mathematical thinking of Voigt, FitzGerald, Lorentz or Poincare, etc., follows Mach´s critical thoughts of absolute simultaneity and omits the ether in the physical discussion. With some practice, then, the conformal transformation for a symmetry group in 4-dimensional space-time is not far away, where the Maxwell equations will remain invariant under the transformation.[xliv] Is in the work of Voigt the wave equation invariant and the speed of light universal, become in Einstein´s theory the laws of physics invariant and the speed of light constant (Ernst and Hsu 2001, p. 225).

## 4. Conclusion

4.1. The relevance of the Doppler effect for the genesis of the special theory of relativity

Many works have been published regarding the genesis of the special theory of relativity. For instance, if we look at the paper by Norton (2004), who has greatly advanced the research on Mach, Einstein, and thought experiments, we see that his reconstruction of the special theory of relativity is different from what I have presented here. In Norton´s reconstruction, for example, the emission theory of light plays an important role (Norton 2004, p. 23). It is certain that Einstein dealt with emission theories, but if and how this influenced him, can be interpreted in different ways and these multiple views result in different reconstructions of the history of science (see Einstein´s thoughts on emission theory in Shankland 1963, pp. 49, 56). Norton examined Einstein´s thought experiment 'chasing the light' and suggests that emission theory could have played a significant role (Norton 2004, pp. 24 f.; Norton 2011, pp. 13 ff.). This thought experiment by Einstein was interpreted differently by some researchers, which are discussed by Norton (2011 pp. 6 ff.). What Norton (2004 & 2011) does not mention is that Susan G. Sterrett suggests that for Einstein´s thought experiment (chasing the light), Mach´s works could have played a role too (Sterrett 1998, pp. 6 f.) and she discusses it in relation to the Doppler effect (Sterrett 1998, p. 25). Moreover, Sterrett refers to another thought experiment by Mach (1878) on the Doppler effect of light (Sterrett 1998, p. 23). In my opinion both authors offer valuable arguments and thus, I would like to argue for a pluralistic reconstruction of the history of science. When scientists try to connect various branches of physics, as was obviously the case with Einstein, they have to research for some time in different direction until they come to a unifying idea. In the history of science too, different sources and interpretations, depending on the focus, can lead to different or to similar results. Seen from this perspective a certain pluralism is unavoidable or even desirable both in historiography (see Kinzel 2016) and in science (see Chang 2012).

Reasons which suggest, that the Doppler effect has had a crucial role in the genesis of the theory of special relativity are briefly listed here.

4.1.1. The special role of the Doppler effect, which appears in both fields (acoustics and optics)

Einstein´s goal was to unify mechanics with electrodynamics. For this purpose, it was necessary to examine the role of the ether more closely and to understand where the analogy between light waves and sound waves led too far. Certainly, by the time Michelson's experiment was published Einstein understood, that the principle of relativity would apply to all inertial systems. The Doppler effect and the aberration of the light could be described without the assumption of an ether. This aberration however, which is related to the Doppler effect, could not be explained by the velocity-addition formula in mechanics. Therefore, he had to consider the speed of light as a constant and redesign the mechanics.

4.1.2. The pioneering role of Woldemar Voigt for the special theory of relativity due to the Doppler effect

Hendrik Antoon Lorentz had already pointed to the pioneering role of Woldemar Voigt in a footnote in his paper on *Elektrogmagnetische Erscheinungen in einem System, das sich mit beliebiger, die des Lichtes nicht erreichender Geschwindigkeit bewegt* (1913) (*Electromagnetic phenomena in a system moving with any velocity smaller than that of light*):

"It is Einstein's credit to be the first person to have verbalized the principle of relativity as a general, strict and precisely valid law.

I add in the remark that Voigt in 1887 (Göttingen Nachrichten p. 41) in his work 'On the Doppler Principle' had already applied a transformation on the following equation

$$\Delta\psi - \frac{1}{c^2}\frac{\partial^2\psi}{\partial t^2} = 0$$

which is contained in equations (4) and (5) of my work. (Note by H.A. Lorentz, 1912)" (Lorentz 1913, p. 10, fn. 1)

A more detailed analysis can be found from Andreas Ernst and Jong-Ping Hsu in their paper *First Proposal of the Universal Speed of Light by Voigt in 1887* (2001). They argue that this work of Voigt "contains several original and fundamental ideas of modern physics". Voigt introduced "the idea of the universal speed of light" and "postulated the invariance of a physics law, the wave equation in 'an elastic incompressible medium'" "to derive the Doppler effect". Furthermore, he analyzed "that the Doppler shift of frequency is incompatible with Newtonian absolute time [...] and is in harmony with a 'relative time'". In addition, Voigt "first derived a type of 4-dimensional space-time transformation" (Ernst and Hsu 2001, pp. 211 f.).

We know today, that Einstein studied Voigt´s considerations early in his carrier (CPAE 1; 126; p. 184). Einstein reconsidered some important concepts in physics in order to arrive at his own theory. We can learn more about this from Einstein's article *The Theory of Relativity* (1915, 1925). Einstein investigated Fizeau´s experiment (CPAE 4; 21; pp. 247 f.) and then studied Lorentz´s physics of electrodynamics (CPAE 4; 21; pp. 248 ff.). (Of course, this coincides with Einstein´s speech.) What goal he pursued, is immediately clear from his summary:

"Let us briefly enumerate the individual results owed so far to the theory of relativity. It yields a simple theory of the Doppler principle, of aberration, of Fizeau's experiment." (CPAE 4; 21; p. 258)

Einstein had always been interested in the experiment of Fizeau in connection with the Doppler effect and the aberration. This becomes apparent in Einstein´s remark on the work of P. Harzers (CPAE 6; 4). In this respect, the section "§7. Theory of Doppler´s principle and aberration" (Einstein 1905, p. 910 ff.) in "On the Electrodynamics of Moving Bodies" was the actual core subject Einstein had taken a critical look at. This is also the reason why section § 7 is located approximately in the middle of Einstein's work.

To sum up the genesis of the theory of special relativity, it may help to begin from the seventh paragraph working our way up to the beginning of the paper[xlv]:

In order to understand the role of the ether, Einstein studied the subject of Doppler effect and the aberration (Einstein 1905, §7; [5]-[16]). This theme has accompanied him all the time. He asked himself how the light would spread in a vacuum (Einstein 1905, §6; [17]-[21]). This in turn led him to reconsider the velocity-addition formula of speed in mechanics (Einstein 1905, §5; [22]-26]). For this, however, he first had to physically interpret the transformation of space and time (Einstein 1905, §1-4; [27]-[42]). (So, if I´m right in my thinking, Einstein did not come up with the findings in his introduction and §1-§2 until the final stages of his research.)

I tried to show that the sections §1 and §2 were developed after a hint from Besso [27] – [39], in which he pointed Einstein to Mach´s concepts (see perhaps [40] – [42]). At that time the introduction, in which Einstein possibly uses Föppl´s thought experiment (without naming him) on a magnet and conductor, was adopted and further developed. This will be discussed in detail later on.

4.2. The reconstruction of the Kyoto speech by Einstein

There are no personal notes from Einstein´s speech in Kyoto. Einstein was accompanied on his trip to Japan by Jun Ishiwara, who was Einstein´s translator. Ishiwara noted the Kyoto speech of Einstein (1922) and published it in Japanese (1923) (Abiko 2000, p. 2). He was a theoretical physicist who had already published works on relativity that were valued by Einstein (Abiko 2000, p. 6). Ishiwara was also the first person to translate and publish the collected scientific works of Einstein in Japanese (1922-1924). Einstein, who in my opinion possessed a great deal of knowledge of human nature, obviously valued and trusted Ishiwara (Abiko 2000, p. 7).

Nevertheless, it is important to review the content of the speech, because Einstein delivered his speech about 17 years after the discovery of the theory of special relativity. For this reason, I have historically reconstructed the content of the speech from the sources (3.1.1-3.1.3). Einstein´s explanation of how he created the theory of relativity is, in my opinion, consistent with the history of science. Some arguments briefly summarized:

(i) The thought experiment that Einstein discusses in his speech, had been previously mentioned in his book quite similarly. There is also evidence in his correspondence that he developed concepts for such an experiment.

(ii) The Michelson experiment, as stated in Einstein´s speech, took place after Einstein´s reflections on his thought experiment.

(iii) Einstein was already familiar with Lorentz´s work prior to the development of the special theory of relativity, he was also deeply involved with Fizeau´s experiment. Evidence of this can be found in various sources.

(iv) Einstein asked and received help from his friend (Besso) developing the theory of special relativity.

Given these reasons, I believe Ishiwara´s notes on Einstein´s Kyoto speech are accurate and an important source for Einstein research. (Of course, the translation of the Japanese text is challenging, but the context is clearly understood from both translations.)

4.3. Historical social background of the Kyoto speech by Einstein

Einstein repeatedly refers to Mach in his Kyoto speech. The reason for this can be seen, above all, if one includes the historical and social background, which was previously overlooked in Einstein research. In section "2. The historical and social background of Einstein´s speech" I examined these areas und put them into context. In doing so, I referred to the Mach-Planck debate to show that Einstein saw himself as a defender of Mach. This is well documented (1) by Einstein's letters to Mach (see Wolters 1987, pp. 148 ff.), (2) by Mach´s endorsement of Einstein's theory of relativity as a progression of his own ideas (Wolters 2019, §28.5), and (3) Einstein´s constant reference to Mach in his work on (general) relativity.

Originally my investigation started off quite differently. I was curious to find out how Einstein reacted to Mach´s preface (he had alleged written). As I explained in the same section, Sommerfeld, a well-known theoretical physicist and supporter of the theory of relativity, drew Einstein´s attention to the newly published foreword by Mach (4 July 1921), in which Mach sees himself as an opponent of the theory of relativity. At that time, Einstein would have had good reason to distance himself from Mach (keyword: Weyland). Therefore, Einstein´s reaction in written form to Sommerfeld is vital – he states the following (13 July 1921):

"It surprises me that Mach did not favor the theory of relativity. For this chain of reasoning lies entirely along his line of thought. I am curious how he arrives at his disavowal." (CPAE 12; 175; p. 125)

Obviously, Einstein saw no contradictions between the theory of relativity and Mach´s considerations. As the founder of the theory of relativity Einstein still views Mach as a forerunner in this field, despite of Mach´s foreword. (Einstein probably did not completely abandon the role of the defender even after the death of Mach.) After having read the correspondence between Sommerfeld and Einstein, it was important to investigate, whether Einstein eventually explained what he meant with "For this chain of reasoning lies entirely along his line of thought." Personally, it was clear to me that he would not have given a detailed written statement on Mach´s foreword. (Obviously, he wanted to wait for the second part – but whether he would have taken a position afterwards, we cannot find out – a second part of optics by Mach was never published.) I then looked into speeches by Einstein around that time and was successful in finding one, in which he referenced Mach. I found out that he was accompanied during his Kyoto speech by Jun Ishiwara, who was a student of Sommerfeld and Einstein (Abiko 2000, p. 6), which also gave me the indication that the content of the speech could be related to the correspondence between Einstein and Sommerfeld. Einstein´s speech "How I created the Theory of Relativity" (1922) can therefore be understood as a kind of defense: he defends the theory of relativity and at the same time Mach´s significant role. It was important to analyze Einstein´s speech more closely to see if there was more evidence on Mach – which led me to my next thesis.

4.4.   Einstein´s thought experiments and their relation to Mach

Given this new perspective (see conclusion 4.3.) it may be a better starting position to re-evaluate Ernst Mach´s influence on Einstein in relation to the emergence of the special theory of relativity. Since Mach can be regarded as one of the pioneers in the theory of thought experiments, I immediately focused on the thought experiments of Einstein. In Einstein´s speech in Kyoto he references two thought experiments. The second thought experiment is concerned with general relativity. This thought experiment, which Einstein called the happiest thought of his life, is known from other sources and has already been investigated by other researchers. It has been shown that Einstein has probably taken this thought experiment from Mach (see Heller 1991, Staley 2013, Staley 2019). (I will elaborate on this in my paper "On thought experiment: Mach and Einstein (Part II)".) Therefore, the question arises whether the first thought experiment, which can be found in part I, is also related to Mach. In this thought experiment Einstein discusses how the relative motion of the Earth towards the ether can be measured with the help of the Doppler effect; and interestingly enough, we can also find a thought experiment from Mach, where Mach discusses the measurability of the relative motion of the Earth towards the Sun. The arguments in favor of Einstein being inspired by Mach´s thought experiment are:

(i)     Einstein reports in a letter to Mileva Maric (10 September 1899):

"A good idea occurred to me in Aarau about a way of investigating how the bodies relative motion with respect to the luminiferous ether affects the velocity of propagation of light in transparent bodies." (CPAE 1; 54; p. 133)

In the same letter he points out that among other things he has already ordered Mach´s works (CPAE 1; 54; p. 133). It is therefore not unlikely that he developed his thoughts with the help of Mach´s work and thus reached the thought experiment, which he discusses in his speech. Finally, both thought experiments use the Doppler Effect to determine the relative motion of the Earth.

(ii) Einstein also mentions this thought experiment in his book *Relativity: The Special and the General Theory*. There are some parallels to Mach´s work in this book: Einstein wrote about this thought experiment (Einstein 2009, p. 35) after a discussion of how the Doppler effect can be measured from spectral lines (Einstein 2009, p. 33). Mach already suggested how to measure the Doppler effect with spectral lines. One page before Mach describes the thought experiment about the Doppler effect, he elaborates on the Machsche Pfeife (Mach´s pipe) (Mach 1873, p. 31). Einstein argues analogously to Mach about the sound of an organ pipe (Einstein 2009, p. 10). If Einstein was concerned with the Doppler effect, then it is obvious that he also read the works of Mach, who he had always appreciated as a physicist.

(iii) Mach points out that his inspiration for his thought experiment came from Fizeau. Einstein will later refer to another experiment by Fizeau as experimentum crucis for special relativity. Mach could have had an influence, among other things, on Einstein to take an in depth look at Fizeau´s experiments. Or Einstein already knew Fizeau´s work pretty well and was therefore interested in Mach´s experiments. Both scenarios are possible.

(iv) Einstein emphasizes (both in his book and in his speech) that in the context of his thought experiment, he no longer thought that one could determine the relative motion of the Earth towards the Sun. This final comment by Einstein is easily understood if it is seen as a response to Mach´s thought experiment. (Of course, other explanations could also be possible.)

The interesting part of Einstein´s thought experiment aren´t necessarily the parallels to Mach´s thought experiment, but particularly the differences. Above all, the changes Einstein made in the thought experiment show what physical problems occupied his mind during this period: is there an absolute movement towards the ether, or are all velocities simply relative, as described in Mach´s works? After having seen the results of the Michelson experiment, he should have recognized the importance of the principle of relativity. Which leads me to my next thesis ...

4.5. Mach´s influence on the theory of special relativity

Based on the assumption that Einstein was inspired by Mach for his thought experiments (see point 4.4.), it was important to investigate if any other concepts of Mach had an influence on the genesis of special relativity. The thought experiment of Mach, which has parallels to Einstein's thought experiment, can be found in his work on the Doppler effect. But was there any additional evidence that Mach´s work on the Doppler effect could have been relevant to the history of special relativity? The clues which suggest this are listed here:

(i) Lorentz, who has dealt both scientifically and historically with the Doppler effect, nominates Mach for the Nobel Prize in Physics (1912). He draws attention to Mach´s performance on acoustics and optics and emphasizes that he had a significant influence on young scientists. (Considering the role of Lorentz in the genesis of special relativity, it is safe to assume that he meant Mach´s influence on Einstein and the theory of relativity amongst others.) Lorentz published a paper entitled *Electromagnetic phenomena in a system moving with any velocity smaller than that of light* – in which he referenced (same year 1912) both Woldemar Viogt´s work *On the Doppler Principle* (1887) and Einstein´s significant achievement (on the principle of relativity) in a footnote. This suggests that Lorentz wanted to emphasize the importance of the Doppler effect for the special theory of relativity.

(ii) Otto Neurath asked for help in a letter to Ernst Mach (1914-1916), on the topic "The light something like the sound", "The sound something like the light". In the same letter, Neurath wrote that he is following the implementation of Mach´s thoughts into the theory of relativity with great interest. After Mach´s death, Neurath would receive a copy of "*The Principles of Physical Optics*" (1921) including the fake preface – his note demonstrates the surprise of this foreword. Neurath may have been the first philosopher of science who understood, the importance of the Doppler effect regarding the genesis of special relativity. He probably did not pursue this idea because of the foreword.

(iii) Max von Laue wrote the following in his review of the book "On Mach´s *The Principles of Physical Optics*" (1921):

"If one were to characterize the situation, it cannot be correct to push scepticism so far that one denies one's own deepest thoughts the moment a person greater than oneself develops it into something positive. And yet the book has rendered a service to relativity theory: He never speaks of a palpable 'Aether' and shows how well all of wave optics can be represented without it." (von Laue 1921, p. 65-66)

In general, the essay shows how disappointed von Laue was about the foreword. Von Laue had given the Doppler effect an important place in his book on the theory of relativity. Unfortunately, I do not know if von Laue was familiar with Mach´s work on the Doppler effect. He may have considered Mach as a forerunner of special relativity, in which Mach´s thoughts were continued by Einstein („that one denies one´s own deepest thoughts the moment a person greater than oneself develops it into something positive").

And finally, he pointed out that Mach´s *The Principles of Physical Optics* had at the very least provided a service to the theory of relativity ("a service to relativity theory") because he disregarded the ether having a special role. One assumption could be that von Laue, as an expert in this field, had knowledge of Mach´s earlier work on the Doppler effect. Even then Mach had tried not to include the ether in the Doppler effect. In this case von Laue´s review could be seen as an indirect reference. To what extent could Einstein have implemented Mach´s thoughts? Which leads me to my next point ...

(iv) (Von Laue could have noticed that) Mach´s argument for the Doppler effect of 1860 does not consider the ether. Mach´s investigations on the Doppler effect in acoustics as well as in optics only take the relative movements between the observer (= receiver) and the source into account. But if one wants to include the propagation medium / ether in their investigations, his concept and the experiment would need to be developed further. The question arises, why Mach did not make these considerations himself? The answer to this question is the same as to why he did not try to reconcile mechanics with the Maxwell equations (to further study the Doppler effect and aberration) ... namely: Mach´s thought experiment is from the year 1862. James C. Maxwell's important work *A dynamical theory of the electromagnetic field*, in which his equations can be found, appeared three years later, in 1865. And as John Norton (Norton 2004, p. 3) already emphasized, electrodynamic is a prerequisite for the special theory of relativity. Einstein seems to point this out in his obituary of Mach:

"It is not improbable that Mach would have hit on relativity theory when in his time - when he was in fresh and youthful spirit-physicists would have been stirred by the question of the meaning of the constancy of the speed of light. In the absence of this stimulation, which flows from Maxwell-Lorentzian electrodynamics, even Mach's critical urge did not suffice to raise a feeling for the need of a definition of simultaneity for spatially distant events." (CPAE 6; 29; p. 144)

Einstein himself clearly identified what was missing in Mach´s work for historical reasons: a theoretical examination of electrodynamics, which would have led him first to the constancy of the speed of light and then to a treatment of simultaneity. It is precisely these stages that Einstein discusses in his speech. (Maybe Max von Laue had considered that?) Since Mach in contrast to Doppler does not investigate the subject of aberration, we can immediately identify what Einstein was looking for: He tried to find out how the Doppler effect and the aberration are related…

(v)  To understand the relativistic aberration, which is associated with the Doppler effect, one needs the relativistic velocity-addition formula. But this contradicted mechanics and Einstein needed help. He had no idea how to physically interpret the "*local time*" ("*Ortszeit*") of Lorentz. As can be seen from Einstein´s speech as well as from the acknowledgment at the end of his work *On the Electrodynamics of Moving Bodies*, a conversation with his friend Besso helped him a lot. From Besso´s letter to Einstein, it can be seen that Besso´s help was generally to point out Mach´s concepts; namely to base physics on the observable. Mach´s critique of absolute space, absolute time and simultaneity should have helped Einstein further ...

(vi) Mach´s book on the Doppler effect and Einstein´s work *On the Electrodynamics of Moving Bodies* show parallels in the manner of argumentation. For example, Mach has been able to solve the discussion between Petzval and Doppler on the basis of various observers. The nature of the argumentation is not surprising, since the Doppler effect can be perceived by observers with relative speeds – for example, when approaching and moving away from trains. The experiments with trains are not only the first example that Mach refers to in his work, but he himself invited professors to Prague (railway station) to verify this fact for themselves. Einstein, for example, in his article *The Theory of Relativity* (1911), which he wrote in his time in Prague, explains the principle of relativity with a thought experiment, in which a physicist is on the train and another is not. On the next page he immediately compared light with sound waves. But that is not the only parallel. He already begins his work *On the Electrodynamics of Moving Bodies* with the critique of simultaneity (Einstein 1905, §1). (Einstein may have been inspired by Mach to criticize simultaneity or absolute time.) Interestingly, Einstein´s first example in this chapter is again a train. Einstein emphasizes that the definition of time depends on the location; or on the relative speed of the observer and how one could synchronize the clocks with the help of the light etc. Since in general the frequency represents the reciprocal of the period time, it is natural that the time has to be transformed. The relativistic Doppler effect thus becomes understandable if one sees the origin of the phenomenon in space and time.

In the next chapter (§2) Einstein will then speak "*On the relativity of lengths and times*". It was important for Einstein to recognize that the relativistic Doppler effect depended only on the frequency of the emitted signal and the relative velocity between A and B. For this reason, he did not care about the ether. In this context it is worth mentioning, that Mach and Einstein address the different perspectives of the observers ("Beobachter"). That Mach uses the term "observer" ("Beobachter") in his explanation in the Doppler effect is understandable, since the Doppler effect directly affects our senses – we can either hear or see it. But it is not self-evident that Einstein in his study of the time uses the term "Beobachter", although it would be sufficient to mention coordinate systems. In § 1 ("*Definition of simultaneity*") he uses the term "Beobachter" three times; in § 2, which discusses with the relativity of lengths and times based on the principle of relativity, he uses the term four times. The next time the term "Beobachter" reappears – as we might expect – is § 7 ("*Theory of Doppler´s principle and of aberration*"). There, Einstein uses the term "Beobachter" six times and thus in his work the most! But this simple concept analysis shows how much the chapters § 1, § 2, and § 7 in Einstein's thought were connected. This can easily be explained if we attribute a special meaning to the Doppler principle in the emergence of the special theory of relativity. It becomes apparent, that Einstein was well acquainted with Mach´s work regarding the Doppler effect.

(vii) Just at the time (1905) when Einstein was engaged in the theory of relativity, Föppl´s paper *On Absolute and Relative Movement* appeared, in which Föppl defended Mach´s thoughts. Perhaps this article encouraged Einstein to further dissect Föppl´s thought experiment (magnet and conductor) on relative motion in electrodynamics, which can be found in the introduction *On the Electrodynamics of Moving Bodies.*

(viii) Susan G. Sterrett suggests, that Mach´s works could have played a role in Einstein´s thought experiment (chasing the light) (Sterrett 1998, pp. 6 f.) and she discusses it in relation to the Doppler effect (Sterrett 1998, p. 25). Moreover, Sterrett refers to another thought experiment by Mach (1878) on the Doppler effect of light (Sterrett 1998, p. 23).

(ix)   Einstein´s friends (Besso and Adler) were Machians. It can also be shown that Einstein intensively studied Mach´s works before he founded the theory of special relativity. It is not unusual, therefore, that he also knew Mach´s work on the Doppler effect, considering the topic occupied him for years.

4.6. On thought experiments: Mach and Einstein

In my analysis, I will rely on an article by Rudolf Haller. This work by Haller, which I consider to be extremely important, is available in both German (Haller 1988) and English (Haller 1992).

According to Rudolf Haller, who interprets Ernst Mach´s reflections in his article *Poetic Imagination and Economy: Ernst Mach as Theorist of Science* the genesis, dynamics and goal of scientific research require two principles: the principle of economics and the poetic imagination (Haller 1988, p. 351 f.) (Haller 1992, p. 224). The principle of economy according to Mach is the principle of distinction between science and non-science (Haller 1988, p. 350) (Haller 1992, p. 223). From facts we try to arrive at laws. The most economical way should be found in order to be able to systematically derive the facts. Economics "is nothing else than the basic hypothesis of the rationality of science, a rationality which is what makes understanding and comprehension possible" (Haller 1988, p. 351) (Haller 1992, p. 224). (Mach´s economy of science is examined well in Lydia Patton´s work *New Water in Old Buckets: Hypothetical and Counterfactual Reasoning in Mach´s Economy of Science* (Patton 2019)). But how can we discover something new from the facts? This is the role of imagination for Mach (Haller 1988, p. 352) (Haller 1992, pp. 224 f.). (I think it is well known that Einstein attributed an importance to imagination, as well.) With the help of creative thinking, we succeed in discovering new structures and thus form new hypotheses and theories. (Rudolf Haller refers to this more precisely: "poetic imagination" (Haller 1988, p.

353) (Haller 1992, p. 226). With our imagination it is also possible to arrive at thought experiments. Haller writes:

"While it is hard to say how to achieve thought-experiments, Mach indicates a way: by means of paradoxes". (Haller 1988, p. 354) (Haller 1992, p. 226)

Paradoxes bring us closer to the nature of a problem. Contradictions aid in keeping our thoughts moving. What does this all have to do with Einstein and the theory of relativity? In the tradition of Mach (Mach 1873, p. 24), Einstein adhered to not view physical terms as a priori (CPAE 6; 29; p. 142). As an experimental physicist, it was important to Mach that physical concepts are always measurable and therefore verifiable and in a relationship to each other. Afterwards one should try to design concepts and theories, which describe the observations as simply as possible, under the principle of economy of thought. In the conversation between Einstein and Heisenberg we can read that Einstein (at a young age in the development of the theory of relativity) represented a specific philosophy (Heisenberg 1969, p. 92; see also Einstein 1949, p. 21), in which physics should be based on observable quantities. This philosophical attitude of building a theory on measurable quantities was in general very important for the emergence of modern physics (quantum mechanics and relativity) (see Feynman, §2.6, p. 2-16). Einstein says that Mach´s economy of thought "probably is part of the truth" for the development of the theory of relativity (Heisenberg 1969, p. 93, my translation). Einstein´s theory of special relativity has been based on "Beobachter" (observer) who can and should measure (relative) lengths and (relative) times. His theory tried to unify mechanics with electrodynamics. That Einstein succeeded with the use of only two principles could be considered an elegant economic act. It was also economical to keep the ether out of the theory. Of course, Einstein needed a lot of imagination to arrive at his two principles from the Doppler principle. That the speed of light in a vacuum is constant, and that length and time can be relative in inertial systems and are dependent on the observer, which is not self-evident or difficult to imagine. But Einstein was able to manage this merely with his thoughts. He was accompanied by a paradox, specifically the contradiction of the velocity addition of mechanics and electrodynamics. The answer apparently came to him after a conversation with his friend Besso, in which he exchanged important thoughts with him.

And in the broadest sense, my work is a "thought experiment" too. Economically, the Doppler principle gets a special role, because it explains why analogies between mechanics and electrodynamics had to be investigated first. The role of the Doppler effect also makes it possible to look at the role of Mach, Voigt, Lorentz and Einstein from a different perspective. In this context, places like Prague (where Christian Doppler, Ernst Mach, Albert Einstein, Philip Frank[xlvi] etc. taught) have a completely different meaning for the history of science. Much of the story fits better together. Of course, this also requires the imagination. There is currently no direct evidence that Einstein knew the works of Mach or Voigt for Doppler effect. (But there are some arguments I have tried to show.) You need some imagination to see these structures. One access to this is provided by the paradox: On the one hand, we have a text (foreword) by Ernst Mach as an opponent of the theory of relativity, which aren´t his own words, and on the other hand, we have Einstein´s speech in Kyoto, which emphasizes Mach as a forerunner, without any written notes from Einstein. The thought of this "paradox" kept me restless. I wanted to give them a voice and I hope that this work was in their favor.

*Acknowledgements:* First and foremost I would like to thank Friedrich Stadler and Bernhard Baumgartner, from whom I have learned a lot about Mach and physics. Their critical comments and support have contributed significantly to this work. I would also like to thank John Norton and James Robert Brown, from whom I gained much knowledge about Mach, Einstein and thought experiments. The discussions with them have been very inspiring to me.

In general, I would also like to thank Elisabeth Nemeth, Gereon Wolters, Gerald Holton, Lydia Patton, Martin Kusch, Franz Embacher, Christoph Dellago, Romano Rupp and Paul Hoffman for their encouraging and critical words. Special thanks go to Hisaki Hashi, who discussed the Japanese text (Kyoto speech) with me. I would like to thank Maria Pflügler, Axel Meckfessel and Btian Paul Dorsam for their support in the translation. Without their help, the translation to English would not be possible, also Valerie Marth for the translation of the French text and her general support. Finally, I would like to thank the reviewers, whose extensive, constructive criticism and hints have taken me very far in my research.

**The Nobel Prize in Physics 1921** [online]: URL: https://www.nobelprize.org/prizes/physics/1921/summary/ [3.6.2019]

**Thiele,** Joachim. 1971. "Zur Wirkungsgeschichte des Dopplerprinzips im Neunzehnten Jahrhundert." *Annals of Science* 27 (4): 393-407

---. 1978. *Wissenschaftliche Kommunikation. Die Korrespondenz Ernst Machs*. Aloys Henn Verlag

**Tipler**, Paul A. and Ralph A. Llewellyn. 2012. Modern Physics. New York: W. H. Freeman and Company

**Treder**, H.-J. 1982. "Der MICHELSON-Versuch als experimentum crucis." Astronomische Nachrichten, 303

**Von Laue**, Max. 1919. *Die Relativitätstheorie. Erster Band: Das Relativitätsprinzip der Lorentztransformation*. Braunschweig: Friedr. Vieweg & Sohn

**Von Laue**, Max. 1921. "On Mach´s *The Principles of Physical Optics*". In *Ernst Mach – a deeper look:* documents and new perspectives, ed. J. Blackmore. Dordrecht: Springer

**Wolters**, Gereon. 1987. *Mach I, Mach II, Einstein und die Relativitätstheorie. Eine Fälschung und ihre Folgen*. Berlin: Walter de Gruyter

---. 1988. "Atome und Relativität – Was meinte Mach?" In *Ernst Mach – Werk und Wirkung,* ed. Rudolf Haller, Friedrich Stadler. Wien: Hölder-Pichler-Tempsky

--- . 2019. "Mach and Relativity Theory: A Neverending Story in HOPOSia?" In *Ernst Mach - Zu Leben, Werk und Wirkung*, ed. F. Stadler. Springer International Publishing

[i] Neither Ishiwara (see Santone 1992, pp. 335 ff.) nor Nishida (see Santone 1992, p. 338) were Machians. (For Nishida´s philosophy of relativity see also Hashi 2007). It is therefore not self-evident that Einstein pointed out Mach in the presence of this audience.

[ii] In the Japanese language, the combination of words is important. These are related to each other in an ('agglutinating') context. A word-for-word translation into a Western language is therefore generally not simple. Also, especially in English, a single word can have multiple meanings. In addition, the Chinese characters in the Japanese written word have particular features, which can lead to different translations.

[iii] In my opinion, the Doppler effect, and in this context the aberration, were themes that always accompanied Einstein´s thinking, during his creation of the special relativity. His interest in emission theory was obvious, among other things, a part from his examination of aberration (see CPAE 6; 7; p. 45 and 49). Generally, there are already simulations that explain the parallels between the Doppler effect and the emission theory. (https://www.einstein-online.info/en/spotlight/doppler/) (For a mathematical and physical introduction see Tipler and Llewellyn 2012, p. 41 f.) I therefore believe that this dilemma is relatively easy to solve.

[iv] Apart from this example, Einstein has never referred to himself as a pupil of another person and certainly never as an admiring student to the addressee. In the same letter Einstein also wrote the following: "I am very happy that the theory of relativity gives you pleasure." (CPAE 5; 175; p. 130)

[v] At the end of Planck's criticism, the following words were directed towards Mach: "[...] unerring mark that divorces false prophets of the true ones: by their fruit you will recognize them!" ("[…] untrügliches Kennzeichen die falschen Propheten von den wahren scheiden lehrt: An ihren Früchten sollt Ihr sie erkennen!") (Planck 1908, p. 51)

As Planck points out, he quotes Jesus at the end (see Matthew 7:20). Gereon Wolters therefore denotes this argument of Planck as "Jesus-Planck-Criterion" ("JESPLAC") (Wolters 2019, § 4) – Personally, I refer to it as the "Sermon on the Mount" of Planck.

[vi] In 1898 he had suffered a stroke and since then was paralyzed on his right side (Haller and Stadler 1988, p. 24)

[vii] Ernst Mach argued (among other things) that Planck did not really understand his thoughts and points out often in his reply (Mach 1910, p. 665, 668-669) and clearly states it in his final sentence:

"But an attempt to understand the opponent should precede it." (my translation)

("Aber ein Versuch, den Gegner zu verstehen, sollte doch vorausgehen.") (Mach 1910, p. 669).

Planck answered that he himself was a follower of Mach´s philosophy (Planck 1910, p. 671). Whether he understood the philosophy of Mach remains questionable (see also Haller and Stadler (1988), pp. 23-24 and 25, footnote 39, 452-454).

[viii] Mach writes to Harald Höffding: "My goal and my job is not to solve philosophical problems, but only to liberate the scientific methodology from old pseudo-problems. What I have tried to give is not a *conclusion*, but a beginning." (Blackmore and Hentschel, p. 44, my translation)

In the original: "Mein Ziel und mein Beruf ist nämlich gar nicht, philosophische Probleme zu lösen, sondern nur die naturwissenschaftliche Methodik von alten im Wege liegenden Pseudoproblemen zu befreien. Was ich zu geben versucht habe ist nicht ein *Abschluss*, sondern ein Anfang." (Blackmore and Hentschel 1985, p. 44)

[ix] A coalition in the sense of mutual support because of (partial and temporary) overlapping of interests, arguments and methods.

[x] The picture (photographed by me) is one of the ten finalists for the competition Arts & Science 2019. (see Arts & Science [online])

[xi] In the original: "Heute wird die Büste E. Machs enthüllt, nicht im Arkadenhofe der Universität, wo die Büsten vieler seiner Berühmten Zeitgenossen stehen, sondern auf einem eigenen Platze, also allein und isoliert."

His article can also be found in Stadler 1988, p. 61. The text "Ernst Mach und die theoretische Physik" from Walter Thirring in which he elaborates on the forged forward can be read also in Stadler 1988, pp.61 f.

[xii] Two examples:

Blackmore, John T. and Klaus Hentschel. 1985. *Ernst Mach als Aussenseiter* ("Ernst Mach the outsider")

Blackmore, John T. 1989. "Ernst Mach Leaves 'The Church of Physics'"

[xiii] The foreword is not from Ernst Mach (Wolters 1987; 2019) and during his lifetime E. Mach has written positively in his publications on the theory of relativity (Wolters 2019). Because the foreword is not from him, but his name is underneath, then, in my opinion, it is a clear case of a fraud. Of course, some questions on this topic are still to be clarified. For an introduction, see Füssl and Hagmann 2017.

[xiv] In *The Collected Papers of Einstein* is a footnote on "(comp. Indroduction)" that says: "See Mach 1921. Published posthumously, the book contains a preface by Mach that may have been written by his son, Ludwig

Mach (see Wolters 1987, but also Holton 1993, who disputes the fabrication)." On the basis of my correspondence (2016) with Gerald Holton I can say that Holton is not against the fake thesis. If there has been a yearlong controversy, it should be considered as resolved.

[xv] After 1945 he fled to the USA. "[H]e worked for the American occupying forces, later for the CIA. On 11 Jan. 1954, Weyland became a citizen of the United States of America and continued there what he had been doing in Germany during the 1920s: he denounced *Einstein* (this time to the American criminal authorities, the FBI). Thus, Weyland returned to the stage where he had started his political career." (Grundmann 2004, p. 109)

[xvi] Einstein writes to Heinrich Zangger (20 January 1914): "The monies for the solar-eclipse undertaking have already been put together in Berlin. Planck behaved very decently in this matter even though he does not believe in the theory, and the whole thing must surely go against his grain because of the Mach polemic. Hats off!" (Beck 1994, p. 377)

[xvii] The whole poem in German and the English translation can be found in Stachel 1982, p. 47 f., fn. 4a. A second English translation can also be found in Abiko 2000, p. 3.

[xviii] "Es ist kein Zweifel, dass das Gedankenexperiment die größten Umwandlungen in unserem Denken einleitet, und die bedeutendsten Forschungswege eröffnet." (Mach 1897, p. 2)

[xix] This structure could disturb some readers. There are three reasons why I would like to refer to this structure:

1. I think the structure shows how Einstein wants to summarize the development of his thoughts. First, he is giving a date. I think it´s normal to start a speech in this way. Next, he explains which concepts he has studied. Then he points to a contradiction that he has tried to solve. In conclusion, he emphasizes that he had the epistemological side also in mind.
2. I believe that this structure can be found in both parts (part I and part II). Since both theories (special and general relativity) was from the same person, it would be interesting if the development of both theories had a common structure.

Didactically, it might be useful to explain the development of the theory of relativity step by step. This structure could possibly contribute to this topic.

[xx] Einstein gives a date when the concept of the theory of special relativity began: 1905 (1922 - 17 = 1905). This means that must have been at least one significant hint at this time, which brought him to an important point after years of searching. I think it was the conversation with his friend Besso, whom he also mentioned in his speech. A second reason could have been the work of Föppl. I will discuss both (Besso and Föppl) in my work and try to explain the relation to Mach.

[xxi] He was interested in optics. What is meant by "optics" can certainly be interpreted differently. Special relativity is a theory to unify mechanics and electrodynamics. For this reason, a topic for the beginning of the research that was already relevant for both fields. If Einstein says optics, then I think, the Doppler effect is, in principle, the most obvious topic.

[xxii] For an introduction on how Einstein's thought experiment on general relativity is related to that of Mach, see Heller 1991, Staley 2013, Staley 2019.

[xxiii] Einstein was strongly influenced by Ernst Mach and Ludwig Boltzmann (see for example (CPAE 1; 54; 133) (CPAE 1; 122; p. 181) (CPAE 1; 136; p. 193). To study Mach and Boltzmann helps to understand also the "miracle year" of Einstein. I will try to explain all this more precisely in my later work.

[xxiv] For an introduction to the topic see for example Thiele (1971) and Hermann (2004).

[xxv] This book is a collection of contributons to the discussion of the Doppler effect by Mach. It includes "Ueber die Aenderungen des Tones und der Farbe durch Bewegung" (1860) ("On the changes of tone and color through Movement") (Mach 1873, pp. 5-21), "Ueber die Controverse zwischen Doppler und Petzval, bezüglich der Aenderung des Tones und der Farbe durch Bewegung" (1861) ("On the controversy between Doppler and Petzval, regarding the change of tone and color by movement") (Mach 1873, pp. 21-29) and "Ueber die Aenderung des Tones und der Farbe durch Bewegung" (1862) ("On the change of tone and color by motion") (Mach 1873, pp. 29-33).

[xxvi] Mach´s defense (1860) of Doppler´s principle can roughly be divided into four areas:

1. Theoretical / mathematical refutation of the critics (Mach 1873, pp. 5-9).

2. Criticism of experimental physicists who believe they have refuted Doppler (Mach 1873, p. 10).

3. Self-built instrument for empirical demonstration of the Doppler effect (Mach 1873, pp. 11-14).

4. Explanation and suggestion on how to make Doppler's principle fruitful for astronomy (Mach 1873, pp. 14-21).

You can clearly see which part of Mach may have been the most significant: the consequences for astrophysics. (He writes eight pages about this theme – half of his work.)

[xxvii] Reference: Exil-Bibliothek Otto Neurath, Wiener Kreis Gesellschaft/Institut Wiener Kreis, Universität Wien, Universitätscampus, Hof 1.

Friedrich Stadler drew my attention to Neurath´s note, for which I am very grateful.

[xxviii] "Interessant ist in diesem Zusammenhang die Stellungnahme des großen Wiener Physikers ERNST MACH. Er, der auch zur Vorgeschichte der Relativitätstheorie Beachtliches beigetragen hat, kam von allen Physikern der wirklichen Bedeutung des DOPPLER-Prinzips am nächsten." (Herrmann 2004, p. 50)

[xxix] For Mach´s pipe see also Mathelitsch and Verovnik (2016) – they start their article with the words: "*The pipe developed by Ernst Mach is not a musical instrument but wrote physics history to the Doppler effect.*" (Mathelitsch and Verovnik 2016, p. 238, my translation)

[xxx] Different stances can be adopted on that interpretation (see Sorensen 1992, pp. 74 f.).

[xxxi] Ernst Mach had different concepts and meanings, when he wrote about thought experiments (TE). Mach describes a specific thought experiment, in which the outcome cannot be known and names it "guessing" ("*Raten*"). He clarifies this with the following words:

"This guessing is not an unscientific process. We can explain this natural process using classic historical examples. A closer look reveals, that this guessing process is responsible for shaping the physical experiment, which is a natural continuation of the thought experiment." (Mach, 1906, p. 194, my translation)

Mach generally distinguishes between raw experience and a planned quantitative experiment (see Mach 1906, p. 193, §8). He then goes into the physical experiments on the next page, etc. Mach therefore differentiates between simple "experience" and a physical "experiment". Buzzoni does not seem to have gone into this differentiation in any more detail (see Buzzoni 2019, p. 653 f.), which could lead to misunderstandings. In general, I think, similar to van Fraassen (see van Fraassen 1980, p. 15), that one should differentiate between *observing* and *observing that* – I think a similar subtle differentiation can also be found in Mach's views. (Unfortunately van Fraassen does not explain or mention the Doppler effect in his important work *The Scientific Image* (1980), which in his philosophical concept would have been a difficult task.)

[xxxii] Mach has not been able to pursue his views because of the strong criticism from his opponents (see Mach 1873, p. 32, fn. 2). The fact that Einstein does not want to refer to the young Mach here might possibly have something to do with the fact that he always saw Mach as a role model in physics and wanted to keep him in the collective memory – a Mach, who had the greatness to criticize Newton skillfully (see CPAE 6; 29; pp. 143 f.).

[xxxiii] The book was written by Einstein in 1916. The first edition appeared in 1917.

[xxxiv] Quite different is it with Christian Doppler. He dedicates whole works to his theory on ether and rotating (celestial) bodies and its dragging effect on light, etc. (see Doppler 2000, pp. 35 ff.).

[xxxv] In his entire work from 1860, the term "ether" occurs only once at the end of the work. At this point he says that "probably" for "stars" that are moving fast "one has to take their resistance in the ether into consideration." (Mach 1873, p. 20, my translation) He thought it is possible that large objects like stars might be slowed down by the ether in their movement. (Perhaps he thought of Doppler's explanations when he wrote this (Doppler 2000, pp. 51, 119 f.). But nowhere Mach associates the ether with the Doppler effect.

[xxxvi] (Mach 1873, p. 8). It should be noted here that Mach did not reject the particle model in principle (in his whole life) (see Wolters 1988) (and even used it at times). Likewise, it is a persistent legend that Ludwig Boltzmann and Mach were enemies because of the atomic theory. Boltzmann saw himself as the successor of Mach (at the University of Vienna) and was in correspondence with him (see Thiele 1978, pp. 44-52).

[xxxvii] In the original:
"Es ist das Verdienst Einsteins, das Relativitätsprinzip zuerst als allgemeines, streng und genau geltendes Gesetz ausgesprochen zu haben.

Ich füge noch in die Bemerkung noch hinzu, daß Voigt bereits im Jahre 1887 (Göttinger Nachrichten S. 41) in einer Arbeit „Über das Dopplersche Prinzip" auf Gleichungen von der Form

$$\Delta \psi - \frac{1}{c^2}\frac{\partial^2 \psi}{\partial t^2} = 0$$

eine Transformation angewandt hat, welche der in den Gleichungen (4) und (5) meiner Arbeit enthalten ist. (Anmerkung von H. A. Lorentz, 1912.)"

[xxxviii] It is interesting that in his epochal and highly influential work "The Structure of Scientific Revolution", where thought experiments are also discussed, Thomas Kuhn does not once mention Mach, even though Kuhn analyzes in detail the Einsteinian revolution (for example Kuhn 1996, pp. 66, 79, 83, 86, 88 f., 98 f., 101 f., 143, 149, 153, etc.).

[xxxix] Einstein´s work was received on June 30. In his speech, he emphasizes that after the conversation with Besso, he needed about five weeks to complete the work.

[xl] In the original:
"Die treffendsten Ausführungen über die physikalische Bedeutung des Trägheitsgesetzes und den damit zusammenhängenden Begriff der absoluten Bewegung rühren von Mach her. Nach ihm ist auch in der Mechanik, wie schon in der Geometrie ohnehin, die Annahme eines absoluten Raumes und hiermit einer absoluten

Bewegung im eigentlichen Sinne unzulässig. Jede Bewegung ist nur als eine relative verständlich und was man gewöhnlich absolute Bewegung nennt, ist lediglich die Bewegung relativ zu einem Bezugssysteme, einem sogenannten Inertialsysteme, das von dem Trägheitsgesetze gefordert wird […]" (Föppl 1904, p. 383)

[xli] Einstein writes in "Fundamental Ideas and Methods of the Theory of Relativity, Presented in Its Development":

"4.⟨The Principle of Special Relativity⟩ The Ether Wind

*One* fatal aspect, however, seemed to cling to Lorentz's theory — namely, the light ether at rest which, more or less, is a materialization of absolute space as Newton introduced it." (CPAE 7; 31; 116)

[xlii] This sequence of steps also reflects the line of argument of Max von Laue: Talking about the "Doppler effect" (von Laue 1919, pp. 26 f.) he immediately discusses the Michelson experiment. For this reason, it is comprehensible that Einstein relates his thought experiment to the Michelson-Morley experiment (von Laue 1919, pp. 27 ff.).

[xliii] "Fundamental Ideas and Methods of the Theory of Relativity, Presented in Its Development":

"The special theory of relativity is nothing but a contradiction-free amalgamation of the results of Maxwell-Lorentz electrodynamics and those of classical mechanics." (CPAE 7; 31; p. 113)

[xliv] For an introduction to the topic see Capria and Manini (2013).

[xlv] At the end of the chapter "§ 8. Transformation der Energie der Lichtstrahlen. Theorie des auf vollkommene Spiegel ausgeübten Strahlungsdruckes." ("§8. *Transformation of the energy of light rays. Theory of the radiation pressure exerted on perfect mirrors*.") Einstein summarizes: "All problems in the optics of moving bodies can be solved by the method employed here." (CPAE 2; 23; p. 165). In Einstein´s speech we can read that he studied "the optics of moving bodies" at the beginning his research to special relativity. In his work in chapter §8 (CPAE 2; 23; p. 165) we find a good indication of what he meant. Of course, this chapter is also related to the Doppler effect. (A brief example at this point may suffice here. In §8 Einstein gives the formula (for the energy of light) $\frac{E'}{E}$ (Einstein 1905, §8, p. 914) (CPAE 2; 23; p. 163). Thereby the inertial system E is at rest and the system E´ moves away from it. (Einstein could have used the formula for the frequency $\frac{f'}{f}$ just as he did in the previous chapter §7 (Einstein 1905, §8, p. 912) (CPAE 2; 23; p. 161), therefore it is unsurprising that §7 (which discusses the Doppler effect) is preceded by §8.

(For this reason, I think it is understandable, that Einstein later examined directly or indirectly the emission theory (E = h · f) in order to better understand the relationship between energy and frequency.)

In any case, the chapter § 9 (and § 10) could be related to [17] - [23] (see for example Einstein 1905, §9, p. 917). Therefore, I think that the topics §9 and §10 were not important for Einstein in the beginning and did not help him discover the special relativity. (Chapters §8-10 are generally more additions.) To arrive at the two principles of special relativity, §1-7 were particularly important.

[xlvi] In a letter Ernst Mach asked Philipp Frank to explain the special relativity and the four-dimensional geometry to him. For this reason Frank visited Mach in Vienna and discussed the theory of relativity with him. A few decades later Frank recounted the events of that meeting and stated that Mach agreed to special relativity, both physically and philosophically (see Wolters 1987, pp. 164 f.).